\documentclass[longauth]{aa}
\usepackage{graphicx}
\usepackage{natbib}
\usepackage{scalerel}
\usepackage{subcaption}
\usepackage[table]{xcolor}
\usepackage{multirow}
\usepackage{breqn}
\usepackage{txfonts}
\usepackage[pdfencoding=auto,psdextra]{hyperref}
\hypersetup{
    colorlinks=true,
    linkcolor=blue,
    filecolor=magenta,      
    urlcolor=blue,
    citecolor=blue
}
\makeatletter
\renewcommand*\aa@pageof{, page \thepage{} of \pageref*{LastPage}}
\makeatother

\usepackage[utf8]{inputenc}

\usepackage[switch, modulo]{lineno}

\usepackage{euclid}

\begin{document}

\title{Euclid Quick Data Release (Q1)}
\subtitle{Hunting for luminous $z > 6$ galaxies in the Euclid Deep Fields -- forecasts and first bright detections}
     
\newcommand{\orcid}[1]{} 
\author{Euclid Collaboration: N.~Allen\orcid{0000-0001-9610-7950}\thanks{\email{natalie.allen@nbi.ku.dk}}\inst{\ref{aff1},\ref{aff2}}
\and P.~A.~Oesch\orcid{0000-0001-5851-6649}\inst{\ref{aff3},\ref{aff1},\ref{aff2}}
\and R.~A.~A.~Bowler\orcid{0000-0003-3917-1678}\inst{\ref{aff4}}
\and S.~Toft\orcid{0000-0003-3631-7176}\inst{\ref{aff1},\ref{aff2}}
\and J.~Matharu\orcid{0000-0002-7547-3385}\inst{\ref{aff5},\ref{aff2}}
\and J.~R.~Weaver\orcid{0000-0003-1614-196X}\inst{\ref{aff6}}
\and C.~J.~R.~McPartland\orcid{0000-0003-0639-025X}\inst{\ref{aff5},\ref{aff2}}
\and M.~Shuntov\orcid{0000-0002-7087-0701}\inst{\ref{aff7},\ref{aff5},\ref{aff2}}
\and D.~B.~Sanders\orcid{0000-0002-1233-9998}\inst{\ref{aff8}}
\and B.~Mobasher\orcid{0000-0001-5846-4404}\inst{\ref{aff9}}
\and H.~J.~McCracken\orcid{0000-0002-9489-7765}\inst{\ref{aff10}}
\and H.~Atek\orcid{0000-0002-7570-0824}\inst{\ref{aff10}}
\and E.~Ba\~nados\orcid{0000-0002-2931-7824}\inst{\ref{aff11}}
\and S.~W.~J.~Barrow\orcid{0009-0001-9835-3650}\inst{\ref{aff1}}
\and S.~Belladitta\orcid{0000-0003-4747-4484}\inst{\ref{aff11},\ref{aff12}}
\and D.~Carollo\orcid{0000-0002-0005-5787}\inst{\ref{aff13}}
\and M.~Castellano\orcid{0000-0001-9875-8263}\inst{\ref{aff14}}
\and C.~J.~Conselice\orcid{0000-0003-1949-7638}\inst{\ref{aff4}}
\and P.~R.~M.~Eisenhardt\inst{\ref{aff15}}
\and Y.~Harikane\orcid{0000-0002-6047-430X}\inst{\ref{aff16}}
\and G.~Murphree\orcid{0009-0007-7266-8914}\inst{\ref{aff8}}
\and M.~Stefanon\orcid{0000-0001-7768-5309}\inst{\ref{aff17}}
\and S.~M.~Wilkins\orcid{0000-0003-3903-6935}\inst{\ref{aff18}}
\and A.~Amara\inst{\ref{aff19}}
\and S.~Andreon\orcid{0000-0002-2041-8784}\inst{\ref{aff20}}
\and N.~Auricchio\orcid{0000-0003-4444-8651}\inst{\ref{aff12}}
\and C.~Baccigalupi\orcid{0000-0002-8211-1630}\inst{\ref{aff21},\ref{aff13},\ref{aff22},\ref{aff23}}
\and M.~Baldi\orcid{0000-0003-4145-1943}\inst{\ref{aff24},\ref{aff12},\ref{aff25}}
\and A.~Balestra\orcid{0000-0002-6967-261X}\inst{\ref{aff26}}
\and S.~Bardelli\orcid{0000-0002-8900-0298}\inst{\ref{aff12}}
\and P.~Battaglia\orcid{0000-0002-7337-5909}\inst{\ref{aff12}}
\and R.~Bender\orcid{0000-0001-7179-0626}\inst{\ref{aff27},\ref{aff28}}
\and A.~Biviano\orcid{0000-0002-0857-0732}\inst{\ref{aff13},\ref{aff21}}
\and E.~Branchini\orcid{0000-0002-0808-6908}\inst{\ref{aff29},\ref{aff30},\ref{aff20}}
\and M.~Brescia\orcid{0000-0001-9506-5680}\inst{\ref{aff31},\ref{aff32}}
\and J.~Brinchmann\orcid{0000-0003-4359-8797}\inst{\ref{aff33},\ref{aff34},\ref{aff35}}
\and S.~Camera\orcid{0000-0003-3399-3574}\inst{\ref{aff36},\ref{aff37},\ref{aff38}}
\and G.~Ca\~nas-Herrera\orcid{0000-0003-2796-2149}\inst{\ref{aff39},\ref{aff40}}
\and V.~Capobianco\orcid{0000-0002-3309-7692}\inst{\ref{aff38}}
\and C.~Carbone\orcid{0000-0003-0125-3563}\inst{\ref{aff41}}
\and J.~Carretero\orcid{0000-0002-3130-0204}\inst{\ref{aff42},\ref{aff43}}
\and G.~Castignani\orcid{0000-0001-6831-0687}\inst{\ref{aff12}}
\and S.~Cavuoti\orcid{0000-0002-3787-4196}\inst{\ref{aff32},\ref{aff44}}
\and K.~C.~Chambers\orcid{0000-0001-6965-7789}\inst{\ref{aff8}}
\and A.~Cimatti\inst{\ref{aff45}}
\and C.~Colodro-Conde\inst{\ref{aff46}}
\and G.~Congedo\orcid{0000-0003-2508-0046}\inst{\ref{aff47}}
\and L.~Conversi\orcid{0000-0002-6710-8476}\inst{\ref{aff48},\ref{aff49}}
\and Y.~Copin\orcid{0000-0002-5317-7518}\inst{\ref{aff50}}
\and F.~Courbin\orcid{0000-0003-0758-6510}\inst{\ref{aff51},\ref{aff52},\ref{aff53}}
\and H.~M.~Courtois\orcid{0000-0003-0509-1776}\inst{\ref{aff54}}
\and M.~Cropper\orcid{0000-0003-4571-9468}\inst{\ref{aff55}}
\and A.~Da~Silva\orcid{0000-0002-6385-1609}\inst{\ref{aff56},\ref{aff57}}
\and H.~Degaudenzi\orcid{0000-0002-5887-6799}\inst{\ref{aff3}}
\and G.~De~Lucia\orcid{0000-0002-6220-9104}\inst{\ref{aff13}}
\and H.~Dole\orcid{0000-0002-9767-3839}\inst{\ref{aff58}}
\and F.~Dubath\orcid{0000-0002-6533-2810}\inst{\ref{aff3}}
\and C.~A.~J.~Duncan\orcid{0009-0003-3573-0791}\inst{\ref{aff47}}
\and X.~Dupac\inst{\ref{aff49}}
\and S.~Dusini\orcid{0000-0002-1128-0664}\inst{\ref{aff59}}
\and S.~Escoffier\orcid{0000-0002-2847-7498}\inst{\ref{aff60}}
\and M.~Farina\orcid{0000-0002-3089-7846}\inst{\ref{aff61}}
\and R.~Farinelli\inst{\ref{aff12}}
\and F.~Faustini\orcid{0000-0001-6274-5145}\inst{\ref{aff14},\ref{aff62}}
\and S.~Ferriol\inst{\ref{aff50}}
\and F.~Finelli\orcid{0000-0002-6694-3269}\inst{\ref{aff12},\ref{aff63}}
\and N.~Fourmanoit\orcid{0009-0005-6816-6925}\inst{\ref{aff60}}
\and M.~Frailis\orcid{0000-0002-7400-2135}\inst{\ref{aff13}}
\and E.~Franceschi\orcid{0000-0002-0585-6591}\inst{\ref{aff12}}
\and M.~Fumana\orcid{0000-0001-6787-5950}\inst{\ref{aff41}}
\and S.~Galeotta\orcid{0000-0002-3748-5115}\inst{\ref{aff13}}
\and K.~George\orcid{0000-0002-1734-8455}\inst{\ref{aff64}}
\and B.~Gillis\orcid{0000-0002-4478-1270}\inst{\ref{aff47}}
\and C.~Giocoli\orcid{0000-0002-9590-7961}\inst{\ref{aff12},\ref{aff25}}
\and J.~Gracia-Carpio\inst{\ref{aff27}}
\and A.~Grazian\orcid{0000-0002-5688-0663}\inst{\ref{aff26}}
\and F.~Grupp\inst{\ref{aff27},\ref{aff28}}
\and S.~V.~H.~Haugan\orcid{0000-0001-9648-7260}\inst{\ref{aff65}}
\and H.~Hoekstra\orcid{0000-0002-0641-3231}\inst{\ref{aff40}}
\and W.~Holmes\inst{\ref{aff15}}
\and I.~M.~Hook\orcid{0000-0002-2960-978X}\inst{\ref{aff66}}
\and F.~Hormuth\inst{\ref{aff67}}
\and A.~Hornstrup\orcid{0000-0002-3363-0936}\inst{\ref{aff68},\ref{aff5}}
\and K.~Jahnke\orcid{0000-0003-3804-2137}\inst{\ref{aff11}}
\and M.~Jhabvala\inst{\ref{aff69}}
\and B.~Joachimi\orcid{0000-0001-7494-1303}\inst{\ref{aff70}}
\and E.~Keih\"anen\orcid{0000-0003-1804-7715}\inst{\ref{aff71}}
\and S.~Kermiche\orcid{0000-0002-0302-5735}\inst{\ref{aff60}}
\and A.~Kiessling\orcid{0000-0002-2590-1273}\inst{\ref{aff15}}
\and B.~Kubik\orcid{0009-0006-5823-4880}\inst{\ref{aff50}}
\and K.~Kuijken\orcid{0000-0002-3827-0175}\inst{\ref{aff40}}
\and M.~K\"ummel\orcid{0000-0003-2791-2117}\inst{\ref{aff28}}
\and M.~Kunz\orcid{0000-0002-3052-7394}\inst{\ref{aff72}}
\and H.~Kurki-Suonio\orcid{0000-0002-4618-3063}\inst{\ref{aff73},\ref{aff74}}
\and A.~M.~C.~Le~Brun\orcid{0000-0002-0936-4594}\inst{\ref{aff75}}
\and D.~Le~Mignant\orcid{0000-0002-5339-5515}\inst{\ref{aff76}}
\and S.~Ligori\orcid{0000-0003-4172-4606}\inst{\ref{aff38}}
\and P.~B.~Lilje\orcid{0000-0003-4324-7794}\inst{\ref{aff65}}
\and V.~Lindholm\orcid{0000-0003-2317-5471}\inst{\ref{aff73},\ref{aff74}}
\and I.~Lloro\orcid{0000-0001-5966-1434}\inst{\ref{aff77}}
\and G.~Mainetti\orcid{0000-0003-2384-2377}\inst{\ref{aff78}}
\and D.~Maino\inst{\ref{aff79},\ref{aff41},\ref{aff80}}
\and E.~Maiorano\orcid{0000-0003-2593-4355}\inst{\ref{aff12}}
\and O.~Mansutti\orcid{0000-0001-5758-4658}\inst{\ref{aff13}}
\and O.~Marggraf\orcid{0000-0001-7242-3852}\inst{\ref{aff81}}
\and M.~Martinelli\orcid{0000-0002-6943-7732}\inst{\ref{aff14},\ref{aff82}}
\and N.~Martinet\orcid{0000-0003-2786-7790}\inst{\ref{aff76}}
\and F.~Marulli\orcid{0000-0002-8850-0303}\inst{\ref{aff83},\ref{aff12},\ref{aff25}}
\and R.~J.~Massey\orcid{0000-0002-6085-3780}\inst{\ref{aff84}}
\and E.~Medinaceli\orcid{0000-0002-4040-7783}\inst{\ref{aff12}}
\and S.~Mei\orcid{0000-0002-2849-559X}\inst{\ref{aff85},\ref{aff86}}
\and Y.~Mellier\inst{\ref{aff7},\ref{aff10}}
\and M.~Meneghetti\orcid{0000-0003-1225-7084}\inst{\ref{aff12},\ref{aff25}}
\and E.~Merlin\orcid{0000-0001-6870-8900}\inst{\ref{aff14}}
\and G.~Meylan\inst{\ref{aff87}}
\and A.~Mora\orcid{0000-0002-1922-8529}\inst{\ref{aff88}}
\and M.~Moresco\orcid{0000-0002-7616-7136}\inst{\ref{aff83},\ref{aff12}}
\and L.~Moscardini\orcid{0000-0002-3473-6716}\inst{\ref{aff83},\ref{aff12},\ref{aff25}}
\and R.~Nakajima\orcid{0009-0009-1213-7040}\inst{\ref{aff81}}
\and C.~Neissner\orcid{0000-0001-8524-4968}\inst{\ref{aff89},\ref{aff43}}
\and S.-M.~Niemi\orcid{0009-0005-0247-0086}\inst{\ref{aff39}}
\and C.~Padilla\orcid{0000-0001-7951-0166}\inst{\ref{aff89}}
\and S.~Paltani\orcid{0000-0002-8108-9179}\inst{\ref{aff3}}
\and F.~Pasian\orcid{0000-0002-4869-3227}\inst{\ref{aff13}}
\and K.~Pedersen\inst{\ref{aff90}}
\and W.~J.~Percival\orcid{0000-0002-0644-5727}\inst{\ref{aff91},\ref{aff92},\ref{aff93}}
\and V.~Pettorino\orcid{0000-0002-4203-9320}\inst{\ref{aff39}}
\and S.~Pires\orcid{0000-0002-0249-2104}\inst{\ref{aff94}}
\and G.~Polenta\orcid{0000-0003-4067-9196}\inst{\ref{aff62}}
\and M.~Poncet\inst{\ref{aff95}}
\and L.~A.~Popa\inst{\ref{aff96}}
\and L.~Pozzetti\orcid{0000-0001-7085-0412}\inst{\ref{aff12}}
\and F.~Raison\orcid{0000-0002-7819-6918}\inst{\ref{aff27}}
\and A.~Renzi\orcid{0000-0001-9856-1970}\inst{\ref{aff97},\ref{aff59}}
\and J.~Rhodes\orcid{0000-0002-4485-8549}\inst{\ref{aff15}}
\and G.~Riccio\inst{\ref{aff32}}
\and E.~Romelli\orcid{0000-0003-3069-9222}\inst{\ref{aff13}}
\and M.~Roncarelli\orcid{0000-0001-9587-7822}\inst{\ref{aff12}}
\and R.~Saglia\orcid{0000-0003-0378-7032}\inst{\ref{aff28},\ref{aff27}}
\and Z.~Sakr\orcid{0000-0002-4823-3757}\inst{\ref{aff98},\ref{aff99},\ref{aff100}}
\and D.~Sapone\orcid{0000-0001-7089-4503}\inst{\ref{aff101}}
\and B.~Sartoris\orcid{0000-0003-1337-5269}\inst{\ref{aff28},\ref{aff13}}
\and M.~Schirmer\orcid{0000-0003-2568-9994}\inst{\ref{aff11}}
\and P.~Schneider\orcid{0000-0001-8561-2679}\inst{\ref{aff81}}
\and T.~Schrabback\orcid{0000-0002-6987-7834}\inst{\ref{aff102}}
\and A.~Secroun\orcid{0000-0003-0505-3710}\inst{\ref{aff60}}
\and E.~Sefusatti\orcid{0000-0003-0473-1567}\inst{\ref{aff13},\ref{aff21},\ref{aff22}}
\and G.~Seidel\orcid{0000-0003-2907-353X}\inst{\ref{aff11}}
\and S.~Serrano\orcid{0000-0002-0211-2861}\inst{\ref{aff103},\ref{aff104},\ref{aff105}}
\and P.~Simon\inst{\ref{aff81}}
\and C.~Sirignano\orcid{0000-0002-0995-7146}\inst{\ref{aff97},\ref{aff59}}
\and G.~Sirri\orcid{0000-0003-2626-2853}\inst{\ref{aff25}}
\and L.~Stanco\orcid{0000-0002-9706-5104}\inst{\ref{aff59}}
\and J.~Steinwagner\orcid{0000-0001-7443-1047}\inst{\ref{aff27}}
\and P.~Tallada-Cresp\'{i}\orcid{0000-0002-1336-8328}\inst{\ref{aff42},\ref{aff43}}
\and A.~N.~Taylor\inst{\ref{aff47}}
\and H.~I.~Teplitz\orcid{0000-0002-7064-5424}\inst{\ref{aff106}}
\and I.~Tereno\orcid{0000-0002-4537-6218}\inst{\ref{aff56},\ref{aff107}}
\and N.~Tessore\orcid{0000-0002-9696-7931}\inst{\ref{aff70},\ref{aff55}}
\and R.~Toledo-Moreo\orcid{0000-0002-2997-4859}\inst{\ref{aff108}}
\and F.~Torradeflot\orcid{0000-0003-1160-1517}\inst{\ref{aff43},\ref{aff42}}
\and I.~Tutusaus\orcid{0000-0002-3199-0399}\inst{\ref{aff105},\ref{aff103},\ref{aff99}}
\and L.~Valenziano\orcid{0000-0002-1170-0104}\inst{\ref{aff12},\ref{aff63}}
\and J.~Valiviita\orcid{0000-0001-6225-3693}\inst{\ref{aff73},\ref{aff74}}
\and T.~Vassallo\orcid{0000-0001-6512-6358}\inst{\ref{aff13},\ref{aff64}}
\and Y.~Wang\orcid{0000-0002-4749-2984}\inst{\ref{aff106}}
\and J.~Weller\orcid{0000-0002-8282-2010}\inst{\ref{aff28},\ref{aff27}}
\and G.~Zamorani\orcid{0000-0002-2318-301X}\inst{\ref{aff12}}
\and F.~M.~Zerbi\inst{\ref{aff20}}
\and E.~Zucca\orcid{0000-0002-5845-8132}\inst{\ref{aff12}}
\and V.~Allevato\orcid{0000-0001-7232-5152}\inst{\ref{aff32}}
\and M.~Ballardini\orcid{0000-0003-4481-3559}\inst{\ref{aff109},\ref{aff110},\ref{aff12}}
\and M.~Bolzonella\orcid{0000-0003-3278-4607}\inst{\ref{aff12}}
\and E.~Bozzo\orcid{0000-0002-8201-1525}\inst{\ref{aff3}}
\and C.~Burigana\orcid{0000-0002-3005-5796}\inst{\ref{aff111},\ref{aff63}}
\and R.~Cabanac\orcid{0000-0001-6679-2600}\inst{\ref{aff99}}
\and M.~Calabrese\orcid{0000-0002-2637-2422}\inst{\ref{aff112},\ref{aff41}}
\and A.~Cappi\inst{\ref{aff12},\ref{aff113}}
\and D.~Di~Ferdinando\inst{\ref{aff25}}
\and J.~A.~Escartin~Vigo\inst{\ref{aff27}}
\and L.~Gabarra\orcid{0000-0002-8486-8856}\inst{\ref{aff114}}
\and W.~G.~Hartley\inst{\ref{aff3}}
\and R.~Maoli\orcid{0000-0002-6065-3025}\inst{\ref{aff115},\ref{aff14}}
\and J.~Mart\'{i}n-Fleitas\orcid{0000-0002-8594-569X}\inst{\ref{aff116}}
\and S.~Matthew\orcid{0000-0001-8448-1697}\inst{\ref{aff47}}
\and M.~Maturi\orcid{0000-0002-3517-2422}\inst{\ref{aff98},\ref{aff117}}
\and N.~Mauri\orcid{0000-0001-8196-1548}\inst{\ref{aff45},\ref{aff25}}
\and R.~B.~Metcalf\orcid{0000-0003-3167-2574}\inst{\ref{aff83},\ref{aff12}}
\and A.~Pezzotta\orcid{0000-0003-0726-2268}\inst{\ref{aff20}}
\and M.~P\"ontinen\orcid{0000-0001-5442-2530}\inst{\ref{aff73}}
\and C.~Porciani\orcid{0000-0002-7797-2508}\inst{\ref{aff81}}
\and I.~Risso\orcid{0000-0003-2525-7761}\inst{\ref{aff20},\ref{aff30}}
\and V.~Scottez\orcid{0009-0008-3864-940X}\inst{\ref{aff7},\ref{aff118}}
\and M.~Sereno\orcid{0000-0003-0302-0325}\inst{\ref{aff12},\ref{aff25}}
\and M.~Tenti\orcid{0000-0002-4254-5901}\inst{\ref{aff25}}
\and M.~Viel\orcid{0000-0002-2642-5707}\inst{\ref{aff21},\ref{aff13},\ref{aff23},\ref{aff22},\ref{aff119}}
\and M.~Wiesmann\orcid{0009-0000-8199-5860}\inst{\ref{aff65}}
\and Y.~Akrami\orcid{0000-0002-2407-7956}\inst{\ref{aff120},\ref{aff121}}
\and I.~T.~Andika\orcid{0000-0001-6102-9526}\inst{\ref{aff122},\ref{aff123}}
\and S.~Anselmi\orcid{0000-0002-3579-9583}\inst{\ref{aff59},\ref{aff97},\ref{aff124}}
\and M.~Archidiacono\orcid{0000-0003-4952-9012}\inst{\ref{aff79},\ref{aff80}}
\and F.~Atrio-Barandela\orcid{0000-0002-2130-2513}\inst{\ref{aff125}}
\and D.~Bertacca\orcid{0000-0002-2490-7139}\inst{\ref{aff97},\ref{aff26},\ref{aff59}}
\and M.~Bethermin\orcid{0000-0002-3915-2015}\inst{\ref{aff126}}
\and A.~Blanchard\orcid{0000-0001-8555-9003}\inst{\ref{aff99}}
\and L.~Blot\orcid{0000-0002-9622-7167}\inst{\ref{aff127},\ref{aff75}}
\and M.~Bonici\orcid{0000-0002-8430-126X}\inst{\ref{aff91},\ref{aff41}}
\and M.~L.~Brown\orcid{0000-0002-0370-8077}\inst{\ref{aff4}}
\and S.~Bruton\orcid{0000-0002-6503-5218}\inst{\ref{aff128}}
\and A.~Calabro\orcid{0000-0003-2536-1614}\inst{\ref{aff14}}
\and B.~Camacho~Quevedo\orcid{0000-0002-8789-4232}\inst{\ref{aff21},\ref{aff23},\ref{aff13}}
\and F.~Caro\inst{\ref{aff14}}
\and C.~S.~Carvalho\inst{\ref{aff107}}
\and T.~Castro\orcid{0000-0002-6292-3228}\inst{\ref{aff13},\ref{aff22},\ref{aff21},\ref{aff119}}
\and F.~Cogato\orcid{0000-0003-4632-6113}\inst{\ref{aff83},\ref{aff12}}
\and S.~Conseil\orcid{0000-0002-3657-4191}\inst{\ref{aff50}}
\and T.~Contini\orcid{0000-0003-0275-938X}\inst{\ref{aff99}}
\and A.~R.~Cooray\orcid{0000-0002-3892-0190}\inst{\ref{aff129}}
\and O.~Cucciati\orcid{0000-0002-9336-7551}\inst{\ref{aff12}}
\and G.~Desprez\orcid{0000-0001-8325-1742}\inst{\ref{aff130}}
\and A.~D\'iaz-S\'anchez\orcid{0000-0003-0748-4768}\inst{\ref{aff131}}
\and J.~J.~Diaz\orcid{0000-0003-2101-1078}\inst{\ref{aff46}}
\and S.~Di~Domizio\orcid{0000-0003-2863-5895}\inst{\ref{aff29},\ref{aff30}}
\and J.~M.~Diego\orcid{0000-0001-9065-3926}\inst{\ref{aff132}}
\and M.~Y.~Elkhashab\orcid{0000-0001-9306-2603}\inst{\ref{aff13},\ref{aff22},\ref{aff133},\ref{aff21}}
\and A.~Enia\orcid{0000-0002-0200-2857}\inst{\ref{aff12}}
\and Y.~Fang\orcid{0000-0002-0334-6950}\inst{\ref{aff28}}
\and A.~G.~Ferrari\orcid{0009-0005-5266-4110}\inst{\ref{aff25}}
\and A.~Finoguenov\orcid{0000-0002-4606-5403}\inst{\ref{aff73}}
\and A.~Fontana\orcid{0000-0003-3820-2823}\inst{\ref{aff14}}
\and F.~Fontanot\orcid{0000-0003-4744-0188}\inst{\ref{aff13},\ref{aff21}}
\and A.~Franco\orcid{0000-0002-4761-366X}\inst{\ref{aff134},\ref{aff135},\ref{aff136}}
\and K.~Ganga\orcid{0000-0001-8159-8208}\inst{\ref{aff85}}
\and J.~Garc\'ia-Bellido\orcid{0000-0002-9370-8360}\inst{\ref{aff120}}
\and T.~Gasparetto\orcid{0000-0002-7913-4866}\inst{\ref{aff14}}
\and V.~Gautard\inst{\ref{aff137}}
\and E.~Gaztanaga\orcid{0000-0001-9632-0815}\inst{\ref{aff105},\ref{aff103},\ref{aff138}}
\and F.~Giacomini\orcid{0000-0002-3129-2814}\inst{\ref{aff25}}
\and F.~Gianotti\orcid{0000-0003-4666-119X}\inst{\ref{aff12}}
\and G.~Gozaliasl\orcid{0000-0002-0236-919X}\inst{\ref{aff139},\ref{aff73}}
\and M.~Guidi\orcid{0000-0001-9408-1101}\inst{\ref{aff24},\ref{aff12}}
\and C.~M.~Gutierrez\orcid{0000-0001-7854-783X}\inst{\ref{aff140}}
\and A.~Hall\orcid{0000-0002-3139-8651}\inst{\ref{aff47}}
\and S.~Hemmati\orcid{0000-0003-2226-5395}\inst{\ref{aff141}}
\and C.~Hern\'andez-Monteagudo\orcid{0000-0001-5471-9166}\inst{\ref{aff142},\ref{aff46}}
\and H.~Hildebrandt\orcid{0000-0002-9814-3338}\inst{\ref{aff143}}
\and J.~Hjorth\orcid{0000-0002-4571-2306}\inst{\ref{aff90}}
\and J.~J.~E.~Kajava\orcid{0000-0002-3010-8333}\inst{\ref{aff144},\ref{aff145}}
\and Y.~Kang\orcid{0009-0000-8588-7250}\inst{\ref{aff3}}
\and V.~Kansal\orcid{0000-0002-4008-6078}\inst{\ref{aff146},\ref{aff147}}
\and D.~Karagiannis\orcid{0000-0002-4927-0816}\inst{\ref{aff109},\ref{aff148}}
\and K.~Kiiveri\inst{\ref{aff71}}
\and J.~Kim\orcid{0000-0003-2776-2761}\inst{\ref{aff114}}
\and C.~C.~Kirkpatrick\inst{\ref{aff71}}
\and S.~Kruk\orcid{0000-0001-8010-8879}\inst{\ref{aff49}}
\and L.~Legrand\orcid{0000-0003-0610-5252}\inst{\ref{aff149},\ref{aff150}}
\and M.~Lembo\orcid{0000-0002-5271-5070}\inst{\ref{aff10},\ref{aff109},\ref{aff110}}
\and F.~Lepori\orcid{0009-0000-5061-7138}\inst{\ref{aff151}}
\and G.~Leroy\orcid{0009-0004-2523-4425}\inst{\ref{aff152},\ref{aff84}}
\and G.~F.~Lesci\orcid{0000-0002-4607-2830}\inst{\ref{aff83},\ref{aff12}}
\and J.~Lesgourgues\orcid{0000-0001-7627-353X}\inst{\ref{aff153}}
\and T.~I.~Liaudat\orcid{0000-0002-9104-314X}\inst{\ref{aff154}}
\and S.~J.~Liu\orcid{0000-0001-7680-2139}\inst{\ref{aff61}}
\and X.~Lopez~Lopez\orcid{0009-0008-5194-5908}\inst{\ref{aff12}}
\and J.~Macias-Perez\orcid{0000-0002-5385-2763}\inst{\ref{aff155}}
\and M.~Magliocchetti\orcid{0000-0001-9158-4838}\inst{\ref{aff61}}
\and F.~Mannucci\orcid{0000-0002-4803-2381}\inst{\ref{aff156}}
\and C.~J.~A.~P.~Martins\orcid{0000-0002-4886-9261}\inst{\ref{aff157},\ref{aff33}}
\and L.~Maurin\orcid{0000-0002-8406-0857}\inst{\ref{aff58}}
\and M.~Miluzio\inst{\ref{aff49},\ref{aff158}}
\and P.~Monaco\orcid{0000-0003-2083-7564}\inst{\ref{aff133},\ref{aff13},\ref{aff22},\ref{aff21}}
\and C.~Moretti\orcid{0000-0003-3314-8936}\inst{\ref{aff13},\ref{aff21},\ref{aff22},\ref{aff23}}
\and G.~Morgante\inst{\ref{aff12}}
\and K.~Naidoo\orcid{0000-0002-9182-1802}\inst{\ref{aff138},\ref{aff70}}
\and A.~Navarro-Alsina\orcid{0000-0002-3173-2592}\inst{\ref{aff81}}
\and S.~Nesseris\orcid{0000-0002-0567-0324}\inst{\ref{aff120}}
\and D.~Paoletti\orcid{0000-0003-4761-6147}\inst{\ref{aff12},\ref{aff63}}
\and F.~Passalacqua\orcid{0000-0002-8606-4093}\inst{\ref{aff97},\ref{aff59}}
\and K.~Paterson\orcid{0000-0001-8340-3486}\inst{\ref{aff11}}
\and L.~Patrizii\inst{\ref{aff25}}
\and R.~Pello\orcid{0000-0003-0858-6109}\inst{\ref{aff76}}
\and A.~Pisani\orcid{0000-0002-6146-4437}\inst{\ref{aff60}}
\and D.~Potter\orcid{0000-0002-0757-5195}\inst{\ref{aff151}}
\and S.~Quai\orcid{0000-0002-0449-8163}\inst{\ref{aff83},\ref{aff12}}
\and M.~Radovich\orcid{0000-0002-3585-866X}\inst{\ref{aff26}}
\and P.-F.~Rocci\inst{\ref{aff58}}
\and G.~Rodighiero\orcid{0000-0002-9415-2296}\inst{\ref{aff97},\ref{aff26}}
\and S.~Sacquegna\orcid{0000-0002-8433-6630}\inst{\ref{aff159}}
\and M.~Sahl\'en\orcid{0000-0003-0973-4804}\inst{\ref{aff160}}
\and E.~Sarpa\orcid{0000-0002-1256-655X}\inst{\ref{aff23},\ref{aff119},\ref{aff22}}
\and A.~Schneider\orcid{0000-0001-7055-8104}\inst{\ref{aff151}}
\and M.~Schultheis\inst{\ref{aff113}}
\and D.~Sciotti\orcid{0009-0008-4519-2620}\inst{\ref{aff14},\ref{aff82}}
\and E.~Sellentin\inst{\ref{aff161},\ref{aff40}}
\and F.~Shankar\orcid{0000-0001-8973-5051}\inst{\ref{aff162}}
\and L.~C.~Smith\orcid{0000-0002-3259-2771}\inst{\ref{aff163}}
\and J.~G.~Sorce\orcid{0000-0002-2307-2432}\inst{\ref{aff164},\ref{aff58}}
\and K.~Tanidis\orcid{0000-0001-9843-5130}\inst{\ref{aff114}}
\and C.~Tao\orcid{0000-0001-7961-8177}\inst{\ref{aff60}}
\and G.~Testera\inst{\ref{aff30}}
\and R.~Teyssier\orcid{0000-0001-7689-0933}\inst{\ref{aff165}}
\and S.~Tosi\orcid{0000-0002-7275-9193}\inst{\ref{aff29},\ref{aff30},\ref{aff20}}
\and A.~Troja\orcid{0000-0003-0239-4595}\inst{\ref{aff97},\ref{aff59}}
\and M.~Tucci\inst{\ref{aff3}}
\and C.~Valieri\inst{\ref{aff25}}
\and A.~Venhola\orcid{0000-0001-6071-4564}\inst{\ref{aff166}}
\and D.~Vergani\orcid{0000-0003-0898-2216}\inst{\ref{aff12}}
\and G.~Verza\orcid{0000-0002-1886-8348}\inst{\ref{aff167}}
\and P.~Vielzeuf\orcid{0000-0003-2035-9339}\inst{\ref{aff60}}
\and N.~A.~Walton\orcid{0000-0003-3983-8778}\inst{\ref{aff163}}
\and D.~Scott\orcid{0000-0002-6878-9840}\inst{\ref{aff168}}}
										   
\institute{Cosmic Dawn Center (DAWN)\label{aff1}
\and
Niels Bohr Institute, University of Copenhagen, Jagtvej 128, 2200 Copenhagen, Denmark\label{aff2}
\and
Department of Astronomy, University of Geneva, ch. d'Ecogia 16, 1290 Versoix, Switzerland\label{aff3}
\and
Jodrell Bank Centre for Astrophysics, Department of Physics and Astronomy, University of Manchester, Oxford Road, Manchester M13 9PL, UK\label{aff4}
\and
Cosmic Dawn Center (DAWN), Denmark\label{aff5}
\and
MIT Kavli Institute for Astrophysics and Space Research, Massachusetts Institute of Technology, Cambridge, MA 02139, USA\label{aff6}
\and
Institut d'Astrophysique de Paris, 98bis Boulevard Arago, 75014, Paris, France\label{aff7}
\and
Institute for Astronomy, University of Hawaii, 2680 Woodlawn Drive, Honolulu, HI 96822, USA\label{aff8}
\and
Physics and Astronomy Department, University of California, 900 University Ave., Riverside, CA 92521, USA\label{aff9}
\and
Institut d'Astrophysique de Paris, UMR 7095, CNRS, and Sorbonne Universit\'e, 98 bis boulevard Arago, 75014 Paris, France\label{aff10}
\and
Max-Planck-Institut f\"ur Astronomie, K\"onigstuhl 17, 69117 Heidelberg, Germany\label{aff11}
\and
INAF-Osservatorio di Astrofisica e Scienza dello Spazio di Bologna, Via Piero Gobetti 93/3, 40129 Bologna, Italy\label{aff12}
\and
INAF-Osservatorio Astronomico di Trieste, Via G. B. Tiepolo 11, 34143 Trieste, Italy\label{aff13}
\and
INAF-Osservatorio Astronomico di Roma, Via Frascati 33, 00078 Monteporzio Catone, Italy\label{aff14}
\and
Jet Propulsion Laboratory, California Institute of Technology, 4800 Oak Grove Drive, Pasadena, CA, 91109, USA\label{aff15}
\and
Institute for Cosmic Ray Research, The University of Tokyo, 5-1-5 Kashiwanoha, Kashiwa, Chiba 277-8582, Japan\label{aff16}
\and
Departament d'Astronomia i Astrof\`isica, Universitat de Val\`encia, C. Dr. Moliner 50, E-46100 Burjassot, Val\`encia, Spain\label{aff17}
\and
Department of Physics \& Astronomy, University of Sussex, Brighton BN1 9QH, UK\label{aff18}
\and
School of Mathematics and Physics, University of Surrey, Guildford, Surrey, GU2 7XH, UK\label{aff19}
\and
INAF-Osservatorio Astronomico di Brera, Via Brera 28, 20122 Milano, Italy\label{aff20}
\and
IFPU, Institute for Fundamental Physics of the Universe, via Beirut 2, 34151 Trieste, Italy\label{aff21}
\and
INFN, Sezione di Trieste, Via Valerio 2, 34127 Trieste TS, Italy\label{aff22}
\and
SISSA, International School for Advanced Studies, Via Bonomea 265, 34136 Trieste TS, Italy\label{aff23}
\and
Dipartimento di Fisica e Astronomia, Universit\`a di Bologna, Via Gobetti 93/2, 40129 Bologna, Italy\label{aff24}
\and
INFN-Sezione di Bologna, Viale Berti Pichat 6/2, 40127 Bologna, Italy\label{aff25}
\and
INAF-Osservatorio Astronomico di Padova, Via dell'Osservatorio 5, 35122 Padova, Italy\label{aff26}
\and
Max Planck Institute for Extraterrestrial Physics, Giessenbachstr. 1, 85748 Garching, Germany\label{aff27}
\and
Universit\"ats-Sternwarte M\"unchen, Fakult\"at f\"ur Physik, Ludwig-Maximilians-Universit\"at M\"unchen, Scheinerstrasse 1, 81679 M\"unchen, Germany\label{aff28}
\and
Dipartimento di Fisica, Universit\`a di Genova, Via Dodecaneso 33, 16146, Genova, Italy\label{aff29}
\and
INFN-Sezione di Genova, Via Dodecaneso 33, 16146, Genova, Italy\label{aff30}
\and
Department of Physics "E. Pancini", University Federico II, Via Cinthia 6, 80126, Napoli, Italy\label{aff31}
\and
INAF-Osservatorio Astronomico di Capodimonte, Via Moiariello 16, 80131 Napoli, Italy\label{aff32}
\and
Instituto de Astrof\'isica e Ci\^encias do Espa\c{c}o, Universidade do Porto, CAUP, Rua das Estrelas, PT4150-762 Porto, Portugal\label{aff33}
\and
Faculdade de Ci\^encias da Universidade do Porto, Rua do Campo de Alegre, 4150-007 Porto, Portugal\label{aff34}
\and
European Southern Observatory, Karl-Schwarzschild-Str.~2, 85748 Garching, Germany\label{aff35}
\and
Dipartimento di Fisica, Universit\`a degli Studi di Torino, Via P. Giuria 1, 10125 Torino, Italy\label{aff36}
\and
INFN-Sezione di Torino, Via P. Giuria 1, 10125 Torino, Italy\label{aff37}
\and
INAF-Osservatorio Astrofisico di Torino, Via Osservatorio 20, 10025 Pino Torinese (TO), Italy\label{aff38}
\and
European Space Agency/ESTEC, Keplerlaan 1, 2201 AZ Noordwijk, The Netherlands\label{aff39}
\and
Leiden Observatory, Leiden University, Einsteinweg 55, 2333 CC Leiden, The Netherlands\label{aff40}
\and
INAF-IASF Milano, Via Alfonso Corti 12, 20133 Milano, Italy\label{aff41}
\and
Centro de Investigaciones Energ\'eticas, Medioambientales y Tecnol\'ogicas (CIEMAT), Avenida Complutense 40, 28040 Madrid, Spain\label{aff42}
\and
Port d'Informaci\'{o} Cient\'{i}fica, Campus UAB, C. Albareda s/n, 08193 Bellaterra (Barcelona), Spain\label{aff43}
\and
INFN section of Naples, Via Cinthia 6, 80126, Napoli, Italy\label{aff44}
\and
Dipartimento di Fisica e Astronomia "Augusto Righi" - Alma Mater Studiorum Universit\`a di Bologna, Viale Berti Pichat 6/2, 40127 Bologna, Italy\label{aff45}
\and
Instituto de Astrof\'{\i}sica de Canarias, E-38205 La Laguna, Tenerife, Spain\label{aff46}
\and
Institute for Astronomy, University of Edinburgh, Royal Observatory, Blackford Hill, Edinburgh EH9 3HJ, UK\label{aff47}
\and
European Space Agency/ESRIN, Largo Galileo Galilei 1, 00044 Frascati, Roma, Italy\label{aff48}
\and
ESAC/ESA, Camino Bajo del Castillo, s/n., Urb. Villafranca del Castillo, 28692 Villanueva de la Ca\~nada, Madrid, Spain\label{aff49}
\and
Universit\'e Claude Bernard Lyon 1, CNRS/IN2P3, IP2I Lyon, UMR 5822, Villeurbanne, F-69100, France\label{aff50}
\and
Institut de Ci\`{e}ncies del Cosmos (ICCUB), Universitat de Barcelona (IEEC-UB), Mart\'{i} i Franqu\`{e}s 1, 08028 Barcelona, Spain\label{aff51}
\and
Instituci\'o Catalana de Recerca i Estudis Avan\c{c}ats (ICREA), Passeig de Llu\'{\i}s Companys 23, 08010 Barcelona, Spain\label{aff52}
\and
Institut de Ciencies de l'Espai (IEEC-CSIC), Campus UAB, Carrer de Can Magrans, s/n Cerdanyola del Vall\'es, 08193 Barcelona, Spain\label{aff53}
\and
UCB Lyon 1, CNRS/IN2P3, IUF, IP2I Lyon, 4 rue Enrico Fermi, 69622 Villeurbanne, France\label{aff54}
\and
Mullard Space Science Laboratory, University College London, Holmbury St Mary, Dorking, Surrey RH5 6NT, UK\label{aff55}
\and
Departamento de F\'isica, Faculdade de Ci\^encias, Universidade de Lisboa, Edif\'icio C8, Campo Grande, PT1749-016 Lisboa, Portugal\label{aff56}
\and
Instituto de Astrof\'isica e Ci\^encias do Espa\c{c}o, Faculdade de Ci\^encias, Universidade de Lisboa, Campo Grande, 1749-016 Lisboa, Portugal\label{aff57}
\and
Universit\'e Paris-Saclay, CNRS, Institut d'astrophysique spatiale, 91405, Orsay, France\label{aff58}
\and
INFN-Padova, Via Marzolo 8, 35131 Padova, Italy\label{aff59}
\and
Aix-Marseille Universit\'e, CNRS/IN2P3, CPPM, Marseille, France\label{aff60}
\and
INAF-Istituto di Astrofisica e Planetologia Spaziali, via del Fosso del Cavaliere, 100, 00100 Roma, Italy\label{aff61}
\and
Space Science Data Center, Italian Space Agency, via del Politecnico snc, 00133 Roma, Italy\label{aff62}
\and
INFN-Bologna, Via Irnerio 46, 40126 Bologna, Italy\label{aff63}
\and
University Observatory, LMU Faculty of Physics, Scheinerstrasse 1, 81679 Munich, Germany\label{aff64}
\and
Institute of Theoretical Astrophysics, University of Oslo, P.O. Box 1029 Blindern, 0315 Oslo, Norway\label{aff65}
\and
Department of Physics, Lancaster University, Lancaster, LA1 4YB, UK\label{aff66}
\and
Felix Hormuth Engineering, Goethestr. 17, 69181 Leimen, Germany\label{aff67}
\and
Technical University of Denmark, Elektrovej 327, 2800 Kgs. Lyngby, Denmark\label{aff68}
\and
NASA Goddard Space Flight Center, Greenbelt, MD 20771, USA\label{aff69}
\and
Department of Physics and Astronomy, University College London, Gower Street, London WC1E 6BT, UK\label{aff70}
\and
Department of Physics and Helsinki Institute of Physics, Gustaf H\"allstr\"omin katu 2, University of Helsinki, 00014 Helsinki, Finland\label{aff71}
\and
Universit\'e de Gen\`eve, D\'epartement de Physique Th\'eorique and Centre for Astroparticle Physics, 24 quai Ernest-Ansermet, CH-1211 Gen\`eve 4, Switzerland\label{aff72}
\and
Department of Physics, P.O. Box 64, University of Helsinki, 00014 Helsinki, Finland\label{aff73}
\and
Helsinki Institute of Physics, Gustaf H{\"a}llstr{\"o}min katu 2, University of Helsinki, 00014 Helsinki, Finland\label{aff74}
\and
Laboratoire d'etude de l'Univers et des phenomenes eXtremes, Observatoire de Paris, Universit\'e PSL, Sorbonne Universit\'e, CNRS, 92190 Meudon, France\label{aff75}
\and
Aix-Marseille Universit\'e, CNRS, CNES, LAM, Marseille, France\label{aff76}
\and
SKAO, Jodrell Bank, Lower Withington, Macclesfield SK11 9FT, UK\label{aff77}
\and
Centre de Calcul de l'IN2P3/CNRS, 21 avenue Pierre de Coubertin 69627 Villeurbanne Cedex, France\label{aff78}
\and
Dipartimento di Fisica "Aldo Pontremoli", Universit\`a degli Studi di Milano, Via Celoria 16, 20133 Milano, Italy\label{aff79}
\and
INFN-Sezione di Milano, Via Celoria 16, 20133 Milano, Italy\label{aff80}
\and
Universit\"at Bonn, Argelander-Institut f\"ur Astronomie, Auf dem H\"ugel 71, 53121 Bonn, Germany\label{aff81}
\and
INFN-Sezione di Roma, Piazzale Aldo Moro, 2 - c/o Dipartimento di Fisica, Edificio G. Marconi, 00185 Roma, Italy\label{aff82}
\and
Dipartimento di Fisica e Astronomia "Augusto Righi" - Alma Mater Studiorum Universit\`a di Bologna, via Piero Gobetti 93/2, 40129 Bologna, Italy\label{aff83}
\and
Department of Physics, Institute for Computational Cosmology, Durham University, South Road, Durham, DH1 3LE, UK\label{aff84}
\and
Universit\'e Paris Cit\'e, CNRS, Astroparticule et Cosmologie, 75013 Paris, France\label{aff85}
\and
CNRS-UCB International Research Laboratory, Centre Pierre Bin\'etruy, IRL2007, CPB-IN2P3, Berkeley, USA\label{aff86}
\and
Institute of Physics, Laboratory of Astrophysics, Ecole Polytechnique F\'ed\'erale de Lausanne (EPFL), Observatoire de Sauverny, 1290 Versoix, Switzerland\label{aff87}
\and
Telespazio UK S.L. for European Space Agency (ESA), Camino bajo del Castillo, s/n, Urbanizacion Villafranca del Castillo, Villanueva de la Ca\~nada, 28692 Madrid, Spain\label{aff88}
\and
Institut de F\'{i}sica d'Altes Energies (IFAE), The Barcelona Institute of Science and Technology, Campus UAB, 08193 Bellaterra (Barcelona), Spain\label{aff89}
\and
DARK, Niels Bohr Institute, University of Copenhagen, Jagtvej 155, 2200 Copenhagen, Denmark\label{aff90}
\and
Waterloo Centre for Astrophysics, University of Waterloo, Waterloo, Ontario N2L 3G1, Canada\label{aff91}
\and
Department of Physics and Astronomy, University of Waterloo, Waterloo, Ontario N2L 3G1, Canada\label{aff92}
\and
Perimeter Institute for Theoretical Physics, Waterloo, Ontario N2L 2Y5, Canada\label{aff93}
\and
Universit\'e Paris-Saclay, Universit\'e Paris Cit\'e, CEA, CNRS, AIM, 91191, Gif-sur-Yvette, France\label{aff94}
\and
Centre National d'Etudes Spatiales -- Centre spatial de Toulouse, 18 avenue Edouard Belin, 31401 Toulouse Cedex 9, France\label{aff95}
\and
Institute of Space Science, Str. Atomistilor, nr. 409 M\u{a}gurele, Ilfov, 077125, Romania\label{aff96}
\and
Dipartimento di Fisica e Astronomia "G. Galilei", Universit\`a di Padova, Via Marzolo 8, 35131 Padova, Italy\label{aff97}
\and
Institut f\"ur Theoretische Physik, University of Heidelberg, Philosophenweg 16, 69120 Heidelberg, Germany\label{aff98}
\and
Institut de Recherche en Astrophysique et Plan\'etologie (IRAP), Universit\'e de Toulouse, CNRS, UPS, CNES, 14 Av. Edouard Belin, 31400 Toulouse, France\label{aff99}
\and
Universit\'e St Joseph; Faculty of Sciences, Beirut, Lebanon\label{aff100}
\and
Departamento de F\'isica, FCFM, Universidad de Chile, Blanco Encalada 2008, Santiago, Chile\label{aff101}
\and
Universit\"at Innsbruck, Institut f\"ur Astro- und Teilchenphysik, Technikerstr. 25/8, 6020 Innsbruck, Austria\label{aff102}
\and
Institut d'Estudis Espacials de Catalunya (IEEC),  Edifici RDIT, Campus UPC, 08860 Castelldefels, Barcelona, Spain\label{aff103}
\and
Satlantis, University Science Park, Sede Bld 48940, Leioa-Bilbao, Spain\label{aff104}
\and
Institute of Space Sciences (ICE, CSIC), Campus UAB, Carrer de Can Magrans, s/n, 08193 Barcelona, Spain\label{aff105}
\and
Infrared Processing and Analysis Center, California Institute of Technology, Pasadena, CA 91125, USA\label{aff106}
\and
Instituto de Astrof\'isica e Ci\^encias do Espa\c{c}o, Faculdade de Ci\^encias, Universidade de Lisboa, Tapada da Ajuda, 1349-018 Lisboa, Portugal\label{aff107}
\and
Universidad Polit\'ecnica de Cartagena, Departamento de Electr\'onica y Tecnolog\'ia de Computadoras,  Plaza del Hospital 1, 30202 Cartagena, Spain\label{aff108}
\and
Dipartimento di Fisica e Scienze della Terra, Universit\`a degli Studi di Ferrara, Via Giuseppe Saragat 1, 44122 Ferrara, Italy\label{aff109}
\and
Istituto Nazionale di Fisica Nucleare, Sezione di Ferrara, Via Giuseppe Saragat 1, 44122 Ferrara, Italy\label{aff110}
\and
INAF, Istituto di Radioastronomia, Via Piero Gobetti 101, 40129 Bologna, Italy\label{aff111}
\and
Astronomical Observatory of the Autonomous Region of the Aosta Valley (OAVdA), Loc. Lignan 39, I-11020, Nus (Aosta Valley), Italy\label{aff112}
\and
Universit\'e C\^{o}te d'Azur, Observatoire de la C\^{o}te d'Azur, CNRS, Laboratoire Lagrange, Bd de l'Observatoire, CS 34229, 06304 Nice cedex 4, France\label{aff113}
\and
Department of Physics, Oxford University, Keble Road, Oxford OX1 3RH, UK\label{aff114}
\and
Dipartimento di Fisica, Sapienza Universit\`a di Roma, Piazzale Aldo Moro 2, 00185 Roma, Italy\label{aff115}
\and
Aurora Technology for European Space Agency (ESA), Camino bajo del Castillo, s/n, Urbanizacion Villafranca del Castillo, Villanueva de la Ca\~nada, 28692 Madrid, Spain\label{aff116}
\and
Zentrum f\"ur Astronomie, Universit\"at Heidelberg, Philosophenweg 12, 69120 Heidelberg, Germany\label{aff117}
\and
ICL, Junia, Universit\'e Catholique de Lille, LITL, 59000 Lille, France\label{aff118}
\and
ICSC - Centro Nazionale di Ricerca in High Performance Computing, Big Data e Quantum Computing, Via Magnanelli 2, Bologna, Italy\label{aff119}
\and
Instituto de F\'isica Te\'orica UAM-CSIC, Campus de Cantoblanco, 28049 Madrid, Spain\label{aff120}
\and
CERCA/ISO, Department of Physics, Case Western Reserve University, 10900 Euclid Avenue, Cleveland, OH 44106, USA\label{aff121}
\and
Technical University of Munich, TUM School of Natural Sciences, Physics Department, James-Franck-Str.~1, 85748 Garching, Germany\label{aff122}
\and
Max-Planck-Institut f\"ur Astrophysik, Karl-Schwarzschild-Str.~1, 85748 Garching, Germany\label{aff123}
\and
Laboratoire Univers et Th\'eorie, Observatoire de Paris, Universit\'e PSL, Universit\'e Paris Cit\'e, CNRS, 92190 Meudon, France\label{aff124}
\and
Departamento de F{\'\i}sica Fundamental. Universidad de Salamanca. Plaza de la Merced s/n. 37008 Salamanca, Spain\label{aff125}
\and
Universit\'e de Strasbourg, CNRS, Observatoire astronomique de Strasbourg, UMR 7550, 67000 Strasbourg, France\label{aff126}
\and
Center for Data-Driven Discovery, Kavli IPMU (WPI), UTIAS, The University of Tokyo, Kashiwa, Chiba 277-8583, Japan\label{aff127}
\and
California Institute of Technology, 1200 E California Blvd, Pasadena, CA 91125, USA\label{aff128}
\and
Department of Physics \& Astronomy, University of California Irvine, Irvine CA 92697, USA\label{aff129}
\and
Kapteyn Astronomical Institute, University of Groningen, PO Box 800, 9700 AV Groningen, The Netherlands\label{aff130}
\and
Departamento F\'isica Aplicada, Universidad Polit\'ecnica de Cartagena, Campus Muralla del Mar, 30202 Cartagena, Murcia, Spain\label{aff131}
\and
Instituto de F\'isica de Cantabria, Edificio Juan Jord\'a, Avenida de los Castros, 39005 Santander, Spain\label{aff132}
\and
Dipartimento di Fisica - Sezione di Astronomia, Universit\`a di Trieste, Via Tiepolo 11, 34131 Trieste, Italy\label{aff133}
\and
INFN, Sezione di Lecce, Via per Arnesano, CP-193, 73100, Lecce, Italy\label{aff134}
\and
Department of Mathematics and Physics E. De Giorgi, University of Salento, Via per Arnesano, CP-I93, 73100, Lecce, Italy\label{aff135}
\and
INAF-Sezione di Lecce, c/o Dipartimento Matematica e Fisica, Via per Arnesano, 73100, Lecce, Italy\label{aff136}
\and
CEA Saclay, DFR/IRFU, Service d'Astrophysique, Bat. 709, 91191 Gif-sur-Yvette, France\label{aff137}
\and
Institute of Cosmology and Gravitation, University of Portsmouth, Portsmouth PO1 3FX, UK\label{aff138}
\and
Department of Computer Science, Aalto University, PO Box 15400, Espoo, FI-00 076, Finland\label{aff139}
\and
Instituto de Astrof\'{\i}sica de Canarias, E-38205 La Laguna; Universidad de La Laguna, Dpto. Astrof\'\i sica, E-38206 La Laguna, Tenerife, Spain\label{aff140}
\and
Caltech/IPAC, 1200 E. California Blvd., Pasadena, CA 91125, USA\label{aff141}
\and
Universidad de La Laguna, Dpto. Astrof\'\i sica, E-38206 La Laguna, Tenerife, Spain\label{aff142}
\and
Ruhr University Bochum, Faculty of Physics and Astronomy, Astronomical Institute (AIRUB), German Centre for Cosmological Lensing (GCCL), 44780 Bochum, Germany\label{aff143}
\and
Department of Physics and Astronomy, Vesilinnantie 5, University of Turku, 20014 Turku, Finland\label{aff144}
\and
Serco for European Space Agency (ESA), Camino bajo del Castillo, s/n, Urbanizacion Villafranca del Castillo, Villanueva de la Ca\~nada, 28692 Madrid, Spain\label{aff145}
\and
ARC Centre of Excellence for Dark Matter Particle Physics, Melbourne, Australia\label{aff146}
\and
Centre for Astrophysics \& Supercomputing, Swinburne University of Technology,  Hawthorn, Victoria 3122, Australia\label{aff147}
\and
Department of Physics and Astronomy, University of the Western Cape, Bellville, Cape Town, 7535, South Africa\label{aff148}
\and
DAMTP, Centre for Mathematical Sciences, Wilberforce Road, Cambridge CB3 0WA, UK\label{aff149}
\and
Kavli Institute for Cosmology Cambridge, Madingley Road, Cambridge, CB3 0HA, UK\label{aff150}
\and
Department of Astrophysics, University of Zurich, Winterthurerstrasse 190, 8057 Zurich, Switzerland\label{aff151}
\and
Department of Physics, Centre for Extragalactic Astronomy, Durham University, South Road, Durham, DH1 3LE, UK\label{aff152}
\and
Institute for Theoretical Particle Physics and Cosmology (TTK), RWTH Aachen University, 52056 Aachen, Germany\label{aff153}
\and
IRFU, CEA, Universit\'e Paris-Saclay 91191 Gif-sur-Yvette Cedex, France\label{aff154}
\and
Univ. Grenoble Alpes, CNRS, Grenoble INP, LPSC-IN2P3, 53, Avenue des Martyrs, 38000, Grenoble, France\label{aff155}
\and
INAF-Osservatorio Astrofisico di Arcetri, Largo E. Fermi 5, 50125, Firenze, Italy\label{aff156}
\and
Centro de Astrof\'{\i}sica da Universidade do Porto, Rua das Estrelas, 4150-762 Porto, Portugal\label{aff157}
\and
HE Space for European Space Agency (ESA), Camino bajo del Castillo, s/n, Urbanizacion Villafranca del Castillo, Villanueva de la Ca\~nada, 28692 Madrid, Spain\label{aff158}
\and
INAF - Osservatorio Astronomico d'Abruzzo, Via Maggini, 64100, Teramo, Italy\label{aff159}
\and
Theoretical astrophysics, Department of Physics and Astronomy, Uppsala University, Box 516, 751 37 Uppsala, Sweden\label{aff160}
\and
Mathematical Institute, University of Leiden, Einsteinweg 55, 2333 CA Leiden, The Netherlands\label{aff161}
\and
School of Physics \& Astronomy, University of Southampton, Highfield Campus, Southampton SO17 1BJ, UK\label{aff162}
\and
Institute of Astronomy, University of Cambridge, Madingley Road, Cambridge CB3 0HA, UK\label{aff163}
\and
Univ. Lille, CNRS, Centrale Lille, UMR 9189 CRIStAL, 59000 Lille, France\label{aff164}
\and
Department of Astrophysical Sciences, Peyton Hall, Princeton University, Princeton, NJ 08544, USA\label{aff165}
\and
Space physics and astronomy research unit, University of Oulu, Pentti Kaiteran katu 1, FI-90014 Oulu, Finland\label{aff166}
\and
Center for Computational Astrophysics, Flatiron Institute, 162 5th Avenue, 10010, New York, NY, USA\label{aff167}
\and
Department of Physics and Astronomy, University of British Columbia, Vancouver, BC V6T 1Z1, Canada\label{aff168}}    
%
%
\abstract{The evolution of the rest-frame ultraviolet luminosity function (UV LF) is a powerful probe of early star formation and galaxy stellar mass build-up. At $z>6$, its bright end ($M_{\rm UV} <  -21$) remains poorly constrained due to small survey volumes of existing near-infrared (NIR) space-based imaging surveys. The Euclid Deep Fields (EDFs) will cover 53\,deg$^2$ with NIR coverage down to 26.5\,AB magnitude, providing a factor of 100 increase in area compared to previous space-based surveys. They thus offer an unprecedented opportunity to select bright $z>6$ Lyman break galaxies (LBGs) and definitively constrain the bright end of the UV LF. With its NIR coverage extending to $\sim2\, \si{\micron}$, \Euclid has the power to detect galaxies out to $z\sim13$. Here, we present a forecast for the number densities of $z>6$ galaxies that \Euclid is expected to observe in the final EDF dataset. Using synthetic photometry from spectral energy distribution (SED) templates of $z=5$--15 galaxies, $z=1$--4 interlopers, and Milky Way MLT dwarfs, we investigate optimal selection methodologies for high-$z$ LBGs in the EDF datasets. We find that a combination of S/N cuts with SED fitting (over optical to MIR bands) yields the highest fidelity sample, recovering more than $76\%$ of the input synthetic $z>6$ LBGs, while limiting low-$z$ contamination to less than $10\%$. This contamination does not include effects from instrumental artefacts, which will impact the first \Euclid data releases. Auxiliary data proves critical: optical coverage from Hyper Suprime Camera and {\it Vera C. Rubin} Observatory will distinguish genuine Lyman breaks from contaminant features, while \Spitzer/IRAC data is vital for recovering $z>10$ sources. Based on empirical double power-law LF models, we expect more than $100\,000$ LBGs at $z=6$--12 and more than $100$ sources as far back as $z>12$ in the final \Euclid data release. In contrast, the steeper Schechter LF models predict no detections of $z>12$ LBGs. In this work, we also present two ultra-luminous ($M_{\rm UV} < -23.5$) candidates selected from the Q1 EDF-N dataset (Euclid Quick Data Release). If their redshifts are reliable, their magnitudes suggest a DPL UV LF model at $z>9$. This highlights the power of \Euclid in constraining the bright end of the UV LF in the early Universe and in identifying the most luminous sources that are valuable for further follow-up observations.}

%
%
%
%
   \titlerunning{Forecasts and first bright detections}
   \authorrunning{Euclid Collaboration: N. Allen et al. }
   
   \maketitle
%
%
%
%
   
\section{\label{sc:EF_Paper_Intro} Introduction}

\begin{figure*}
    \centering
    \includegraphics[width=1\linewidth]{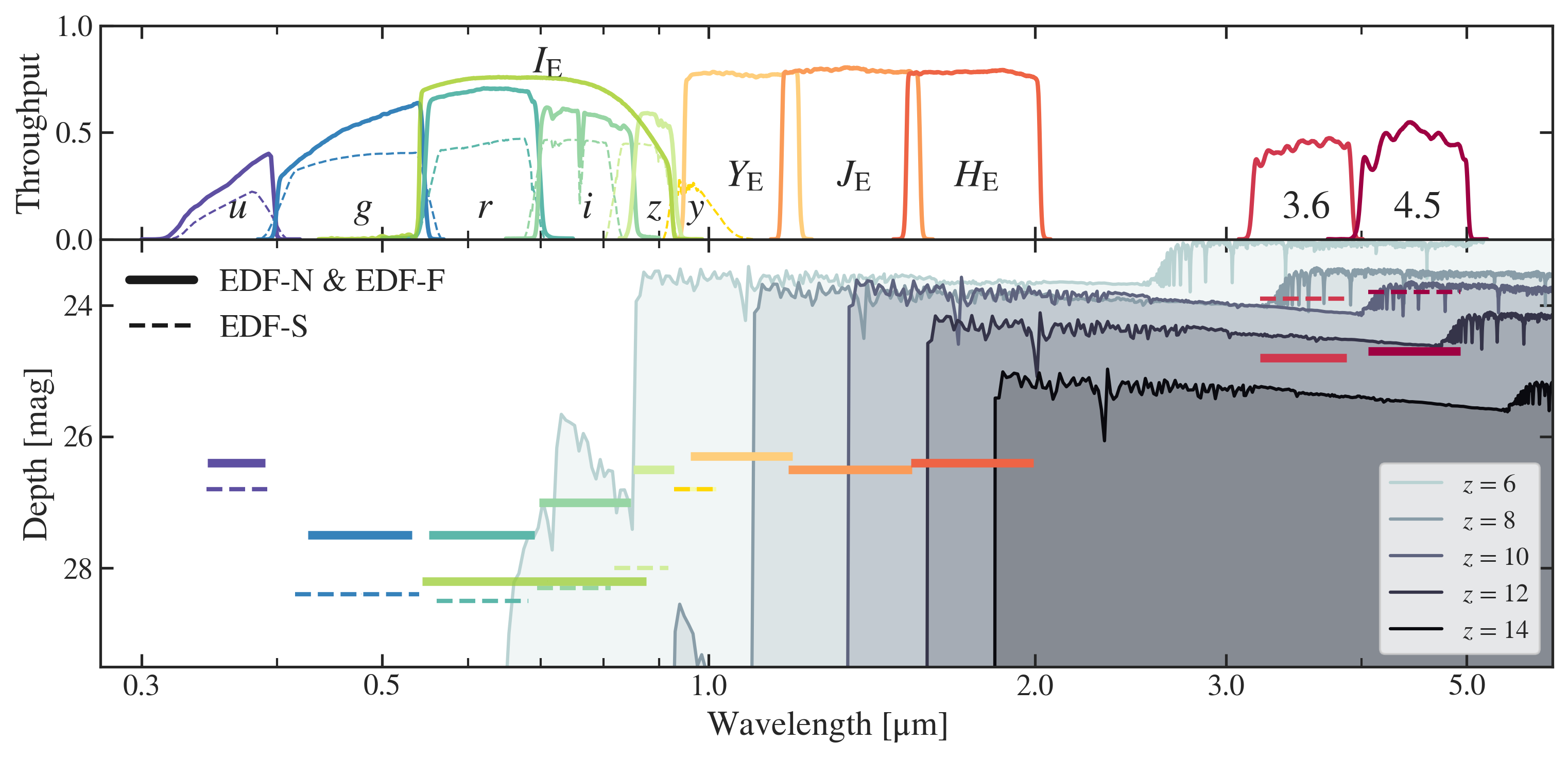}
    \caption{\textit{Top panel}: The throughput curves for CFHT/MegaCam, Subaru/HSC, \Euclid and {\it Spitzer}/IRAC filters as solid lines and Rubin/LSST filters as dashed lines. \textit{Bottom Panel}: The depths of each filter (5$\sigma$), presented as solid lines for EDF-N and EDF-F and dashed lines for EDF-S. {\it Spitzer}/IRAC depths are average measurements from the images from \cite{Moneti-EP17}. The depths of \Euclid VIS \citep{EuclidSkyVIS} and NISP (\citealt{EuclidSkyNISP, Schirmer-EP18}) filters are those expected for point sources by the end of the mission. Expected depths for Rubin data from \citet{foley2018}. Subaru HSC and CFHT MegaCam depths for COSMOS are from \citet{COSMOS2020}. Flexible stellar population synthesis template \citep{FSPS-1,FSPS-2} examples of LBGs at $z= 6,8,10,12$, and 14 shown in blue/grey colours.
    The NIR filters of \Euclid can detect galaxies out to $z\sim12$--14, before the Ly$\alpha$ break shifts out of the \HE band. }
    \label{fig:sed-filters-lbgs}
\end{figure*} 

In the first few hundred million years of the Universe's history, under- and over-dense regions of dark matter formed in the cosmic web. The latter are believed to have seeded the first galaxies, which merged and evolved into massive systems at later times \citep{White&Reese1978, Mo2010_GalaxyFormationEvolution}. In the current picture of galaxy evolution, these systems often host centrally luminous sources, surrounded by fainter companions that contribute to ionising the surrounding intergalactic medium, IGM \citep{Mo2010_GalaxyFormationEvolution}. 
Identifying and characterising the most luminous sources in the early Universe is crucial for understanding early galaxy formation, their physical properties (e.g., star-formation rates), their role in cosmic reionisation, and their connection to the underlying dark matter distribution. 
However, early over-dense regions are rare and require wide and deep surveys to be identified \citep[e.g.][]{bowler2020_uvlf_ultravista, bowler2015, Ono18_uv_lf, harikane_goldrish_uvlf, Casey24_CWz10}.

The rest-frame ultraviolet (UV) luminosity function (LF) is a key observable for tracing galaxy evolution, as the shape and evolution of the LF parameters give insights into galaxy growth and star-formation mechanisms (e.g., \citealt{Silk&Mamon12_GalaxyReview, Mutch13_UVLF}). 
Using combined optical/NIR data from the Wide Field Camera 3 (WFC3) on the {\it Hubble} Space Telescope (HST) and MIR data from \Spitzer -IRAC, the rest-frame UV LF of galaxies has been measured out to $z=11$ based on Lyman break galaxies (LBGs; \citealt{Steidel&Gamilton1993_LBG}). LBGs are UV bright, star-forming galaxies with strong breaks at 1216$\,\AA$ (at $z>6$) caused by absorption of their UV emission by neutral hydrogen in the IGM (see e.g. \citealt{Madau95,Inoue2014}). The shape of the UV LF is still debated: studies selecting LBGs at $z>6$ with HST and {\it Spitzer} found that the Schechter function \citep{schechter}  was an optimal model for the high-$z$ UV LF (\citealt{Ellis_2013, mclure_2013, Finkelstein_2015, mcloed_2015, mcleod_dunlop_2016, bouwens2021_uvlf}). 
The Schechter function is derived from the shape of the halo mass function \citep{press-schechter} with a few modifications. The function drops off exponentially at the bright end, suggesting that the growth of high-mass star-forming galaxies is hindered either by active galactic nuclei (AGN) heating, inefficient gas cooling due to the heat from shocks, and/or attenuation from dust (c.f. \citealt{harikane_goldrish_uvlf, bowler2020_uvlf_ultravista, Ono18_uv_lf}; and see \citealt{Stark2025_review} for reviews). 

However, studies using larger area surveys with ground-based telescopes, e.g. UltraVISTA \citep{McCraken_ULTRAVISTA}, Subaru/Hyper Suprime Camera (HSC), or UKIRT \citep{Lawrence_UKIRT}, found an excess of bright $z>6$ sources, in comparison to previous Schechter function fits to a fainter sample. This suggested a change in evolution at the bright end at $z>6$. A double power-law model (DPL) was suggested to describe the UV LF at these high redshifts (e.g. \citealt{Bowler2014_z7UVLF, Ono18_uv_lf,
bowler2020_uvlf_ultravista, harikane_goldrish_uvlf, Kauffmann22_COSMOS2020_UVLF, varadaraj23_uvlf}). The DPL shows a more gradual decline at the bright end, indicating a lack of dust attenuation or mass quenching by, e.g., inefficient feedback from AGN or Supernova (\citealt{Bowler2017}; see \citealt{stark2016_review} for a review).

With the launch of the {\it James Webb} Space Telescope (JWST), the high-$z$ Universe (at $z>10$) has become accessible, allowing us to constrain the formation of the first galaxies ($z\ge 10$). Studies using JWST data have measured the high-$z$ LF up to $z\sim 13$ and found a surprisingly large number density of luminous ($M_{\rm UV} > -21$) galaxies at $z>8$; up to 10 times higher than expected based on previous measurements from HST (e.g. \citealt{Naidu22a,donnan24_jwst_uvlf, McLeod23_JWSTUVLF, Adams23_UVLF, bouwens23_jwst_uvlf, whitler25_jades_uvlf, robertson24_jades_uvlf, Castellano22_JWSTBrightz9}; and see \citealt{Adamo24_review} for a review). These measurements suggest that the growth mechanisms for early galaxies may be changing at $z\gtrsim9$ and pose a challenge for theoretical models of galaxy evolution (\citealt{Mason2023_HighSFRs, ferrara24,  feldmann25, dekel23, yung24, Lovell2021_Flares, Vijayan2021_Flares, hutter24}). Subsequently, theoretical models have attempted to find the mechanism which would produce such a bright population of early galaxies. Some of the possible solutions that have been proposed are: low attenuation from dust (due to radiative winds: \citealt{ferrara24}), bursty star-formation (\citealt{geli24_bursty-sf, Mason2023_HighSFRs}), an evolving IMF (\citealt{hutter24}), an increase in star-formation efficiency, or reduced feedback (\citealt{dekel23, Mason2023_TheoreticalBrightz10,Somerville25}).

Within this context, the \Euclid mission is perfectly positioned to resolve the debate about the shape of the UV LF during the epoch of reionisation. 
\Euclid will observe $\rm \sim 53\, deg^2$ as part of the Euclid Deep Fields (EDFs) reaching $5\sigma$ depths of $\rm 26.5 AB$ in NIR, by the end of the 6-year mission \citep{EuclidSkyOverview}. This is approximately two orders of magnitude larger in area than any existing space-based imaging from HST or JWST, allowing us to constrain the bright end of the UV LF up to $z\sim 13$ and reduce cosmic variance. 
However, at these depths, ultra-cool dwarf number densities (e.g., M, L, and T types) peak, which could result in significant contamination in high-$z$ selections due to similar red colours in optical and NIR bands (\citealt{Wilkins2014_BD, Bowler2014_z7UVLF, varadaraj23_uvlf}. When observed in broadband filters, the spectral features due to molecular absorption in MLT atmospheres mimic the Lyman break and make these populations difficult to remove in colour selection criteria. Removing MLTs with \Euclid bands alone has been shown by \citet{Banados25} to be difficult. To reduce contamination in photometric samples, it is thus important to use a wide wavelength coverage, from optical to MIR (\citealt{bowler2015, varadaraj23_uvlf}).

In this paper, we predict the expected yield of $z>6$ LBGs that can be identified with \Euclid over the 6-year survey duration, based on the latest estimates of the UV LF evolution from JWST. We then present selection criteria that minimise contamination rates of low-$z$ interlopers and ultra-cool dwarfs to less than $10\%$, while keeping the recovery rates of $z>6$ sources above $70\%$. The synthetic catalogue used in this work contains simulated photometry from Subaru/HSC, Rubin/LSST (Large Synoptic Survey Telescope), \Euclid, and {\it Spitzer} filters and assumes the final 6-year mission coverage regarding the area and depth of all these facilities. 

This paper is structured as follows: Sect. \ref{sc:EF_Paper_Data} describes the mock catalogues used for the predictions, and Sect. \ref{sc:EF_Paper_Method} describes the optimised selection methods. We present our results in Sect. \ref{sc:EF_Paper_Results}, followed by discussion in Sect. \ref{sc:EF_Paper_Discussion}. Examples of ultra-bright Q1 candidates are presented in Sect. \ref{sc:EF_Paper_Discussion_Q1Examples}. The work in this paper is summarised in the conclusion in Sect. \ref{sc:EF_Paper_Conclusion}. Throughout this paper, we use a standard flat, cold dark matter cosmology with $\Omega_{\rm m} = 0.27$, $\Omega_\Lambda = 0.73$, and $H_0 = 70\, \mathrm{km s}^{-1} \mathrm{Mpc}^{-1}$. 
The magnitudes used in this paper are specified in the AB system \citep{ok31974_ABmag} and our SED fitting assumes a \citet{chabrier_imf} initial mass function (IMF).

\section{\label{sc:EF_Paper_Data}Synthetic Euclid Deep Field catalogues}

\begin{figure}
    \centering
    \includegraphics[width=1\linewidth]{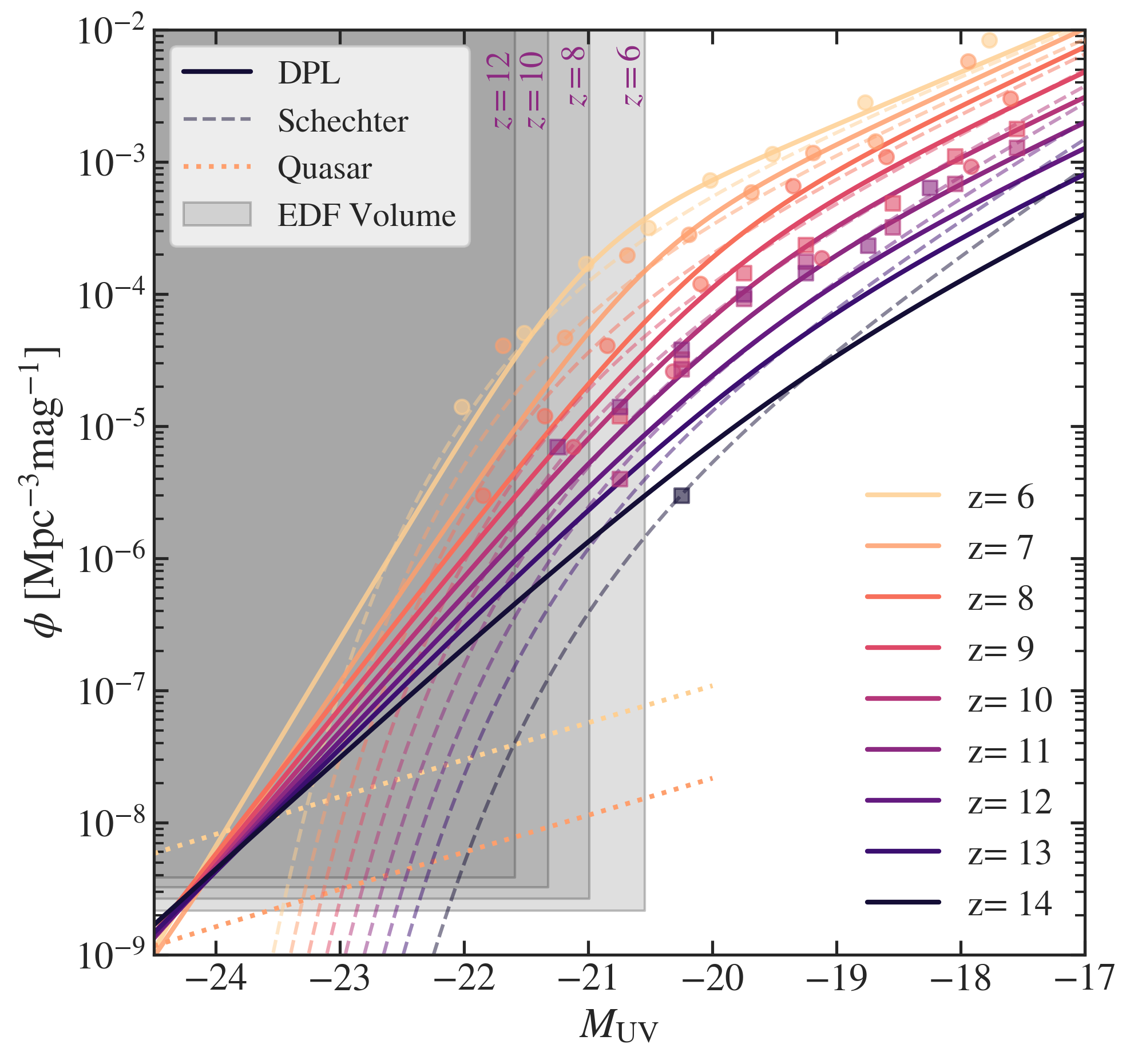}
    \caption{Evolution of the input UV luminosity function from $z=6$ to 14 with parametrised evolutions presented in Eqs. (\ref{equ:bowler_LF_param_ev}) to (\ref{equ:donnan_LF_param_ev}). These LF models are taken from a JWST-based study \citep{donnan24_jwst_uvlf} between $8\le z < 15$ and a HST and ground-based study \citep{bowler2020_uvlf_ultravista} between $6\le z < 8$. These parametrisations are used for the number densities of high-$z$ sources in our sample (see Sect. \ref{sc:EF_Paper_Data_highz}). Circle and square markers are data points from \citet{donnan24_jwst_uvlf} at $z>9$ and \citet{bouwens2021_uvlf} at $z<9$. The shaded regions indicate the volume and depth achieved by the EDFs at the end of the mission (at various redshifts). Dashed lines show an alternative parametrisation of the UV LF using a Schechter function (\citealt{whitler25_jades_uvlf}). Dotted lines show the fiducial LF model from \citet{Schindler23_QuasarLF} for quasars at $z=5$ and $z=6$. With the final \Euclid data release, we will be able to distinguish between these models.}
    \label{fig:input_lf}
\end{figure}
This paper aims to investigate the number of high-$z$ ($z>6$) galaxies that \Euclid will yield and to test how well these sources can be recovered, including possible contamination. The high-$z$ LBGs ($z>6$) are characterised by a strong break in their SEDs at the redshifted Ly$\alpha$ line, in addition to blue colours longward of the break (see Fig. \ref{fig:sed-filters-lbgs}). However, these characteristics can be mimicked by ultra-cool dwarfs in the Milky Way, and quiescent or dusty galaxies at lower redshift \citep[see also][]{vanMierlo-EP21}. To test for possible contamination by these populations, we create synthetic catalogues that include these three types of sources with the expected number densities for the three EDFs. We include all relevant filters for each field that will be available, including Subaru HSC, LSST, \Euclid, and {\it Spitzer} IRAC bands, and perturb the photometry with Gaussian random scatter that mimics the uncertainties in the real data. 

In the following, we first briefly discuss the characteristics of the different EDFs before describing the input number densities and SED shapes that were used for each of the three simulated populations.

\subsection{Simulated fields}
For this paper, we generate synthetic catalogues for the three EDFs separately, using the respective areas and datasets that will be available by the end of the mission. Specifically, we simulate the EDF-N field that covers the North Ecliptic Pole with an area of 20 deg$^2$, the EDF-S with an area of 23 deg$^2$ near the Southern Ecliptic Pole, and the EDF-F with 10 deg$^2$ in the Fornax constellation near the Chandra Deep Field South \citep[see][for details]{Scaramella-EP1,EP-McPartland}. 

The different fields will be covered with different ancillary data by the end of the mission. The field with the most extensive multi-wavelength coverage will be EDF-F, where optical data are obtained with the CFHT, Subaru telescopes, and the upcoming {\it Vera C. Rubin} Observatory (VRO) covering the $u$- to $y$-band filters.  
In contrast, the EDF-N field will not be covered by Rubin/LSST but will have Subaru optical coverage \citep{Capak2016_SpitzerSubaruProposal}. EDF-S will contain Rubin/LSST, but not Subaru/HSC or CFHT/MegaCam.
Importantly, all fields have coverage with relatively deep {\it Spitzer}/IRAC imaging in the 3.6 and 4.5 $\micron$ bands (see \citealt{Moneti-EP17, Capak2016_SpitzerSubaruProposal}), although this is limited to only 10 $\rm deg^2$ of {\it Spitzer}/IRAC in EDF-N. 

For each field, we generate synthetic catalogues with fluxes perturbed with Gaussian noise. Specifically for this idealised simulation, we convert the expected 5$\sigma$ depths ($m_5$) for a given filter into a 1$\sigma$ flux uncertainty $\sigma_{\rm f}$, which is used to perturb the photometry of galaxies, i.e. the measurement uncertainties are set to be constant for each filter with a value of
{$\sigma_{\rm f} = 0.2\times 10^{-0.4(m_5-\mathrm{ZP})}$}, where $\mathrm{ZP}$ is the AB magnitude zero-point of the given filter. We note that these errors may be underestimated because we have not simulated other sources of error, e.g. the effects of artefacts, in the real \Euclid images.
A full lists of the available filters and depths are listed in Table 2 in \citet{EP-McPartland}.

\begin{figure*}
    \centering
    \includegraphics[width=1\linewidth]{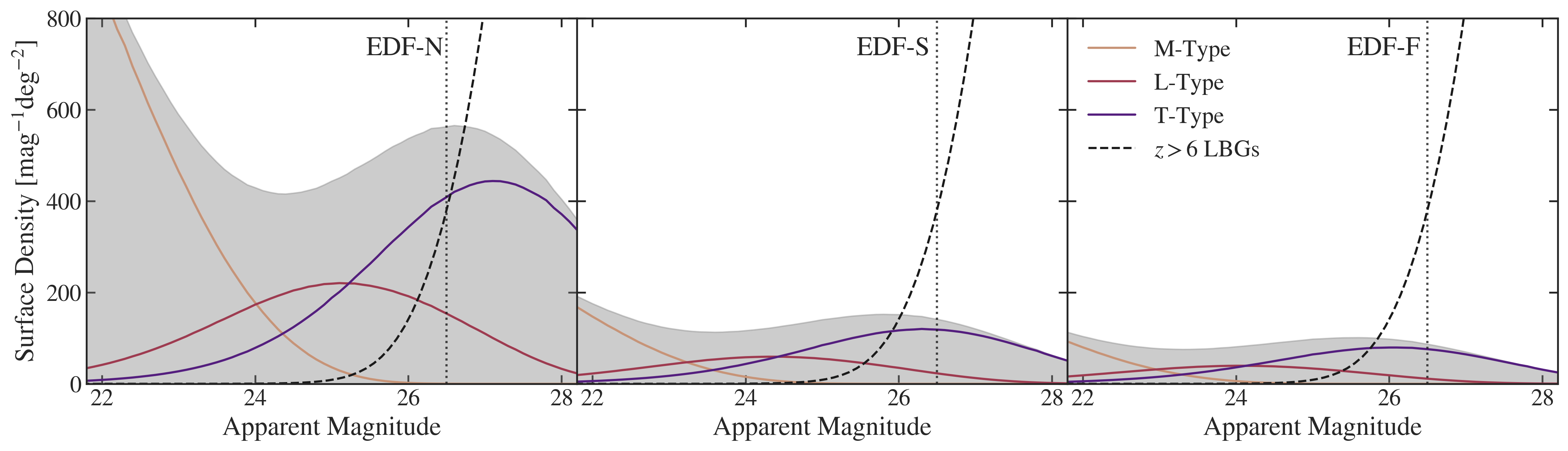}
    \caption{The simulated number densities of M (tan), L (red), and T (purple) dwarfs at different (apparent) magnitudes, for each EDF. The grey shaded region shows the sum of all MLT number densities. The number counts of LBGs at $z>6$ are shown as the black dashed line and the final 6-year mission depth of the EDFs as a black dotted line. The number densities of MLT dwarfs are largest in EDF-N at all magnitudes, which contributes to larger contamination rates.}
    \label{fig:browndwarf_nomdens_per-field}
\end{figure*}
\subsection{\label{sc:EF_Paper_Data_highz}High-redshift LBGs}

Recent JWST observations have revealed a surprisingly large number density of $M_{\rm UV} < -22$ galaxies at $z>8$; up to 10 times higher than expected based on previous measurements from HST (e.g. \citealt{bouwens2021_uvlf}). Additionally, the most recent observations covering wide areas also indicate that the UV LF of $z\sim5$--8 galaxies does not behave like a Schechter function with an exponential cut-off at the bright end but rather exhibits a double power-law shape (e.g., \citealt{bowler2015, harikane_goldrish_uvlf, bowler2020_uvlf_ultravista, donnan24_jwst_uvlf}). This means that we can now expect \Euclid to detect galaxies even beyond $z\sim10$, which was not the case with HST-based extrapolations of the UV LFs. 

The expected numbers of $z>5$ galaxies in the EDFs are derived based on the latest estimates of the UV LFs from JWST at $z=6$--14, as well as wide-area HSC and UltraVISTA measurements, which are available up to $z\sim7$. We use a DPL LF with parameters that evolve with redshifts according to estimates from
\citet{bowler2020_uvlf_ultravista} at $z\le 8$ and \citet{donnan24_jwst_uvlf} at $z \ge 8$. These are chosen due to these studies combining multiple surveys to increase the area probed at high-$z$. The evolution of the luminosity functions is presented in Fig. \ref{fig:input_lf}. 

In terms of absolute UV magnitude, the DPL LF is parametrised as

\begin{equation}
    \phi(M) = \frac{\mathrm{d}n}{\mathrm{d}M} = \frac{\phi_*}{10^{-0.4 (M^*_{\rm UV} - M_{\rm UV})(\alpha +1)} + 10^{-0.4 (M^*_{\rm UV} - M_{\rm UV})(\beta +1)}},
\end{equation} 
with $\phi_*$ defined as the characteristic density, $M^*$ the characteristic magnitude, and $\alpha$ and $\beta$ the faint- and bright-end slopes, respectively.

The evolution of the parameters used for the $5<z<8$ LFs is based on \citet{bowler2020_uvlf_ultravista} and are defined as
\begin{gather}
\label{equ:bowler_LF_param_ev}
    \log_{10} \left(\frac{\phi_*(z)}{\rm mag^{-1}\,Mpc^{-3}}\right) = -3.52, \\
    M^*(z) = -21.03 + 0.49\,(z - 6.0), \\
    \alpha(z) = -1.99 - 0.09\,(z - 6.0), \\
    \beta(z) = -4.92 + 0.45\,(z - 6.0).
\end{gather}

For LFs beyond $z=8$, we follow \citet{donnan24_jwst_uvlf}, who measured the LFs at $8\leq z \leq 15$ in various JWST fields and found the following redshift evolution 
\begin{gather}
\label{equ:donnan_LF_param_ev}
    \log_{10} \left(\frac{\phi_*(z)}{\rm mag^{-1}\,Mpc^{-3}}\right) = -0.14\, (z-2.36), \\
    M^*(z) = -20.95 + 0.11\,z , \\
    \alpha(z) = -2.04 \times 10^{-4}\,z - 2.1, \\
    \beta(z) = 0.138\,z - 5.13 .
\end{gather}

The combination of these two parametrisations results in a smooth evolution of the UV LFs both at the bright and faint ends, which agrees with the latest measurements across the full luminosity and redshift range, as shown in Fig. \ref{fig:input_lf}. 

In the following, we use this DPL LF evolution as our baseline model. However, we will also discuss a parametrisation of the LF using a Schechter function. In particular, we use the estimated evolution of Schechter function parameters from $z=5$ to $z=14$ in \citet{whitler25_jades_uvlf}, which combined HST and JWST data. We note again that JWST only probes survey volumes that are less than 10$^6$\, Mpc$^3$ and hence only has limited power to distinguish between the two LF models.

Based on these LFs, we compute the expected number of galaxies as a function of UV absolute magnitude, $M_{\rm UV}$, for a given survey area and per redshift through integration. Galaxies are added to our catalogue in redshift bins from $z=5$ to $z=15$ in steps of $\Delta z=0.1$. 

For each galaxy, we then assign an SED. These are generated from \citet{BC03} models with a constant star-formation history, an age ranging from 10 to 400 Myr, and a range of UV attenuation following \citet{Calzetti2001_DustAttenuation}. The reddening values are chosen such that the UV continuum slope distribution of galaxies follows a luminosity and redshift-dependent trend in agreement with recent measurements from JWST, in particular \citet{Topping2024_UVSlopes}. We use a linear relation between the UV slope, $\beta_{\rm UV}$ and $M_{\rm UV}$, parametrized as
\begin{equation}
\beta_{\rm UV}(M_{\rm UV}) = \frac{\mathrm{d}\beta_{\rm UV}}{\mathrm{d}M_{\rm UV}}\, (M_{\rm UV}+19) + \beta_{\rm UV,0} + N(0,\sigma_{\rm \beta,int}), 
\end{equation}
where the redshift-dependent slopes and intercepts are from \citet{Topping2024_UVSlopes}, which are listed in Table \ref{tbl:beta-muv}. We interpolate linearly between the redshifts where the slope and intercept measurements are provided. The Gaussian intrinsic scatter of $N(0,\sigma_{\rm \beta,int})$ around the mean relation was set to be constant $\sigma_{\rm \beta,int}=0.25$, consistent with \citet{Topping2024_UVSlopes} and previous HST-based measurements \citep[see e.g.,][]{bouwens2014_UVConSlopes}.

The rest-frame optical emission lines and the nebular continuum are added to these templates using the prescription of \citet{AndersFritze03} and \citet{Schaerer2002}. Finally, absorption by intergalactic hydrogen shortward of the redshifted Ly$\alpha$ line is accounted for following \citet{Inoue2014}. These templates are then used to compute the expected fluxes in all the relevant filters and normalised to the $M_{\rm UV}$ values of the galaxies to be simulated before Gaussian photometric scatter is applied, as described in the previous section.

\begin{table}
\centering
\caption{The redshift dependent parameters of the UV slope $\beta_{\rm UV}$ -- luminosity relation used in this work.}
\label{tbl:beta-muv}
\setlength{\tabcolsep}{0.5em} 
{\renewcommand{\arraystretch}{1.3}
\begin{tabular}{c|c|c}
$\langle\,z\,\rangle$  & $\frac{\mathrm{d}\beta_{\rm UV}}{\mathrm{d}M_{\rm UV}}$ & $\beta_{\rm UV,0}$ \\
\hline  \hline 
5.86 & $-0.11$ & $-2.25$ \\
7.28 & $-0.12$ & $-2.26$ \\
9.41 & $-0.06$ & $-2.33$ \\
12.02 & $-0.06$ & $-2.42$ \\
\end{tabular}}
\end{table}

\begin{figure*}
    \centering
    \includegraphics[width=1\linewidth]{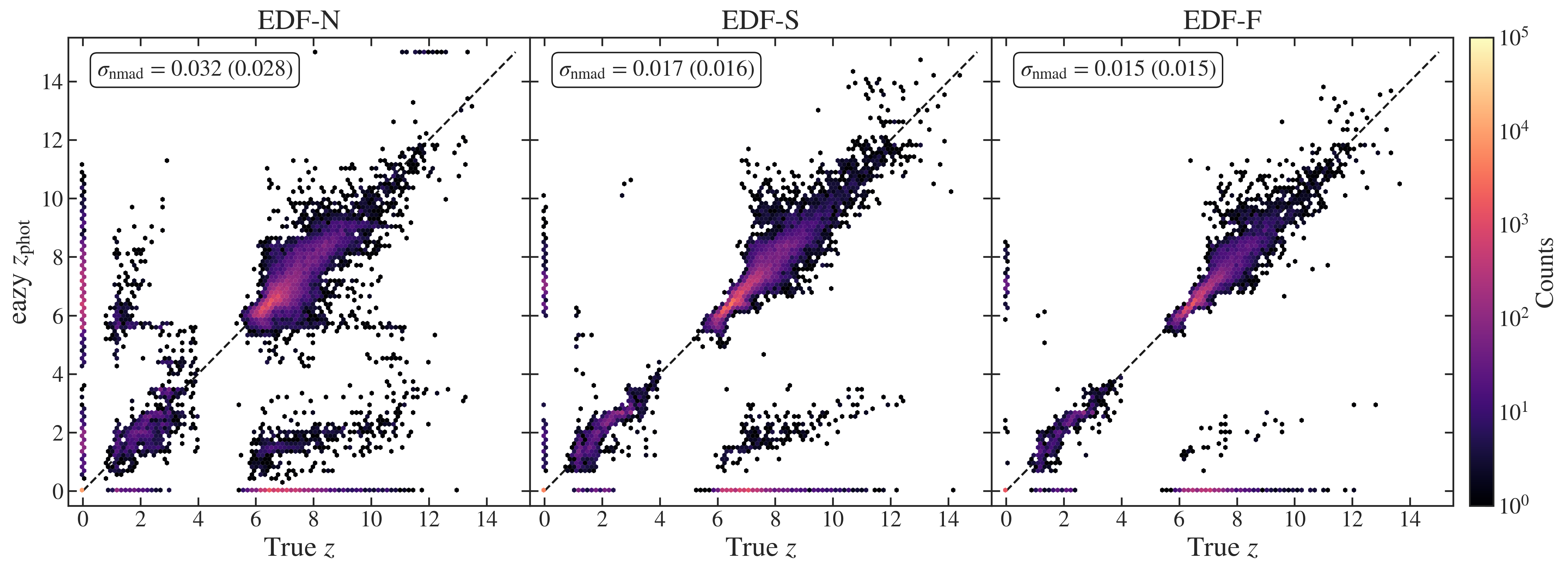}
    \caption{Hexbin distribution of the \texttt{eazy} best-fit photometric redshifts of the input catalogue against their true redshift, for each EDF. Note that MLTs are placed at $z=0$. The scatter in the distribution, $\sigma_{\rm nmad}$, is shown in the top left corner of each panel. The scatter is largest in EDF-N due to shallower optical data and the lack of {\it Spitzer}/IRAC coverage in half the field.}
    \label{fig:EAZY_all_fields_zphot_zpec}
\end{figure*}

\subsection{\label{sc:EF_Paper_Data_browndwarfs}MLT dwarfs}

Certain types of stars or ultra-cool dwarfs (e.g. M-, L-, and T-type) are a possible source of contamination when selecting $z>6$ galaxies. We will refer to these as MLTs. They have intrinsically red optical to NIR molecular absorption features, which can mimic the Lyman break of $z>6$ LBGs, especially close to the magnitude limit or in the case of low numbers of red-optical and NIR filters that poorly sample the shape of the SED. 
MLTs can be removed based on their point source morphology, but this requires high-resolution imaging (e.g. JWST resolution). Although the \Euclid NIR bands have higher resolution than ground-based telescopes, morphology may not be a reliable method to distinguish between stars and high-$z$ sources, especially for compact high-$z$ sources.
Their number densities also peak around $m\sim 24$\,AB \citep{hst-browndwarf_numdens}, which indicates that they might be a significant contaminant for \Euclid datasets. 

We have included synthetic photometry of MLTs that reproduce \Euclid colour - colour tracks derived in \citet{sanghi24_bd_color} for types from M0 to T9. Since our goal is also to simulate the IRAC photometry of MLTs, we cannot rely on the BD standards in the commonly used SpexPrism library~\citep{Burgasser2014}, which are limited in wavelength coverage to the {\it K} band. 
Instead, we use a range of templates that reproduce the predicted \Euclid colour - colour tracks selecting from the BT-Settl, CIFIST, and ATMO2020 libraries (CIFIST2011/2015; \citealt{Allard2012, Baraffe2015, Phillips2020}, as also used in \citealt{sanghi24_bd_color}).

The MLT number densities for each EDF are computed based on a simple exponential model of the Milky Way thin disk (e.g. \citealp{hst-browndwarf_numdens, Carnero2019, Holwerda2014, Holwerda2024}). 
The number densities thus depend on the Galactic latitude of a given field. 
We follow the simple thin disk parametrisation of~\citet{Caballero2008} to predict the number of MLTs within each field, using $h_R = 2250\,{\rm pc}$, $R_\odot = 8.6\,{\rm kpc}$, and $Z_{\odot} = 27\,{\rm pc}$ as the radial scale height, the Solar radius and the Solar vertical displacement respectively. 
The exact number density in our parametrisation depends on the local space density and the vertical-scale height ($h_{\rm Z}$) assumed.
We take the values from~\citet{Caballero2008} for the local space density.
The vertical scale height is uncertain with values derived in the range 150--450\,pc (e.g.~\citealp{Sorahana2019}).
In this work, we fix $h_{\mathrm{Z}} = 300\,{\rm pc}$ to provide a first estimate of the expected contamination.

As can be seen in Fig. \ref{fig:browndwarf_nomdens_per-field}, the expected number densities for M dwarfs decrease to fainter \JE-band magnitudes. However, the number densities of L and T dwarfs together can be as high as 400\,$\rm mag^{-1}deg^{-2}$ at the \JE-band detection limit of the EDFs, making them comparable to the expected densities of high-redshift LBGs. As shown in Fig. \ref{fig:browndwarf_nomdens_per-field}, EDF-N will have the highest contamination rate due to the field being close to the Galactic plane. 
While not all these sources will necessarily fall within the high-$z$ selection performed on \Euclid data, the comparable colours and number densities make them a key contamination concern.
Hence, it will be important to be able to remove these sources efficiently from high-$z$ candidate catalogues.

\subsection{\label{sc:EF_Paper_Data_lowz} Low-$z$ interlopers}

Low-redshift (low-$z$) sources can also have colours that mimic a Lyman break, especially if they have a strong Balmer break or emission lines that boost the NIR filters (\citealt{Naidu22, Carnall23_ExtremeEmissionLines, Haro23}). The low-$z$ contamination in the EDFs has already been extensively studied in \citet{vanMierlo-EP21}. Thus, in this work, we only add a minimal set of possible lower redshift SEDs. In particular, we only add sources with $ i-J> 1$. These were selected from 3D-HST catalogues \citep{Skelton2014} directly using their \texttt{eaz} \citep{Brammer08_eazy} template fits. The synthetic colours of the lower redshift red galaxy population were computed based on the best-fit \texttt{eaz} template at the respective photometric redshift before adding Gaussian random noise to perturb the photometry. Because the 3D-HST survey only spans $\sim900$ arcmin$^2$, each red galaxy SED is represented several times in the synthetic catalogues, to mimic the larger areas of the EDFs. 

\section{\label{sc:EF_Paper_Method} High-redshift galaxy selection criteria}
With three simulated catalogues for each EDF in hand, we now discuss the optimal selection method which returns the highest completeness levels of high-$z$ sources while minimising the contamination rate. Other selection criteria that we experimented with, but that are less optimal, are discussed in the Appendix \ref{sc:EF_Paper_Discussion_OtherCrietria}.

The initial catalogues were created with fainter magnitude limits than can be observed with \Euclid, to allow for scattering into the sample due to noise. Therefore, the first cut we employ is a detection criterion based on signal-to-noise ratio (S/N).  We apply a $\mathrm{S/N}\ge 5$ in at least one of the \Euclid NISP bands (\YE, \JE, or \HE) giving us a parent sample size of $1\,256\,398$, $1\,444\,968$, and $613\,682$ in EDF-N, EDF-S, and EDF-F respectively, including all three types of sources we simulated. To identify a clean high-$z$ sample, we proceed as follows. 

Due to the characteristic break at rest-frame $\lambda = 1216\,\AA$ in high-$z$ LBGs, we do not expect to measure significant signal blueward of the Lyman break. Therefore, we apply a $\rm S/N < 2$ cut to the \IE band to remove sources with optical flux. With this cut, we can remove $\sim 99\%$ and $\sim 55\%$ of low-$z$ sources and MLTs from the parent sample, respectively, while keeping the recovery rate of $>70\%$ for $z>6$ LBGs, in all EDFs. Applying similar S/N cuts in the ground-based bands can reduce the MLT contamination by a few percent. However, these cuts remove more high-$z$ galaxies ($\sim 10\%$) compared to the removal of MLTs. This is because of the up-scattering in optical bands and some residual, but real signal from the Lyman alpha forest in $z=6$ galaxies. Therefore, we do not use ground-based data in the initial cuts and instead remove MLTs with SED fitting (this is discussed further in the Appendix \ref{sc:EF_Paper_Discussion_Others_ext-opt}).

The final sample is selected based on photometric redshifts ($z_{\rm phot}$) measured with SED fitting with \texttt{eaz}. To optimise the fitting of high-$z$ sources, we use the \texttt{sfhz/blue\_sfhz\_13.param} stellar population templates for the galaxy fitting. We set the \texttt{eaz} parameters to fit in the range $z=0$--15 (with $\Delta z = 0.01$), no priors, and IGM optical depth of $\tau_{\rm IGM}$ of 1. We do not consider damped Lyman-$\alpha$ templates. To simulate a real fitting process, we do not supply \texttt{eaz} with the known redshifts of each source. 
All other parameters are unchanged and set to the default values supplied in the \texttt{eaz} default parameter file. 

We set \texttt{eaz} to fit templates across all bands in each EDF, including ground-based instruments (Subaru/HSC or/and Rubin/LSST), \Euclid bands (VIS and NISP), and {\it Spitzer} IRAC (Channel 1 and 2). This large wavelength coverage is important for reliable SED fitting and to remove contamination; see Sect. \ref{sc:EF_Paper_Discussion_Others_NoOptExt} for details.
To remove MLTs from our final sample, we also fit MLT templates based on the Sonora library \citep{Marley2018_SoronaTemplate}. We then compare the reduced $\chi^2$ of the galaxy and stellar template fits to remove likely MLTs. The final high-$z$ candidate sample is selected using the following cuts

\begin{enumerate}
    \item $z_{\rm phot} \ge 5$,
    \item $\chi^2_{\rm star} > \chi^2_{\rm galaxy}$.
\end{enumerate}

Here, $z_{\rm phot}$ is the best-fit photometric redshift measured by \texttt{eaz}. The $\chi^2_{\rm star}$ and $\chi^2_{\rm galaxy}$ are the minimised $\chi^2$ values measured from the best-fit stellar and galaxy templates, respectively.

The results of the SED fitting in each field are shown in Fig. \ref{fig:EAZY_all_fields_zphot_zpec}. For all fields, the scatter between the photometric and true redshifts of the input sample is low, $\sigma_{\rm nmad} \le 0.032$. Here, $\sigma_{\rm nmad}$ is the standard deviation using the median absolute deviation of $z_{\rm true}-z_{\rm phot}$. This shows that we can fit reliable SED templates and, thus, reliable photometric redshifts. We discuss this further in Sect. \ref{sc:EF_Paper_Results_RecoveryPurity}. 

Our method selects all sources that have a best-fit solution at $z\ge 6$, which allows for sources with second peaks in their $p(z)$ in the final sample. We estimate the probability of having a second peak, $p_{\rm 2nd}$, by integrating over the $p(z)$ up to $z=5$ of each true high-$z$ source (that is classified as high-$z$). We then assume that sources with $p_{\rm 2nd} > 0.05$ have a second peak. Therefore, the fraction of true high-$z$ sources that are classified as high-$z$, which have a second peak, is $18\%$, $6\%$, and $4\%$ for EDF-N, EDF-S, and EDF-F, respectively

\begin{figure*}
     \centering
     \begin{subfigure}{\textwidth}
         \centering
         \includegraphics[width=0.78\textwidth]{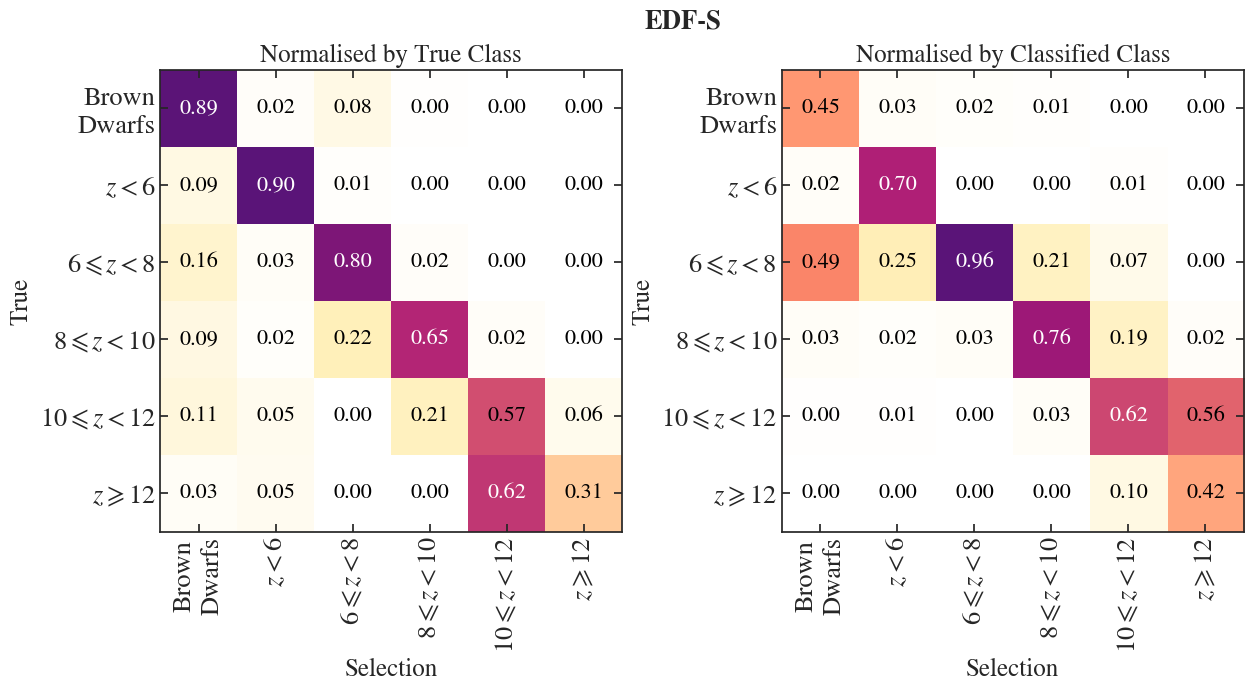}
         \label{fig:recoved_sel_frac_sep}
     \end{subfigure}
     \hfill
     \begin{subfigure}{\textwidth}
         \centering
         \includegraphics[width=0.78\textwidth]{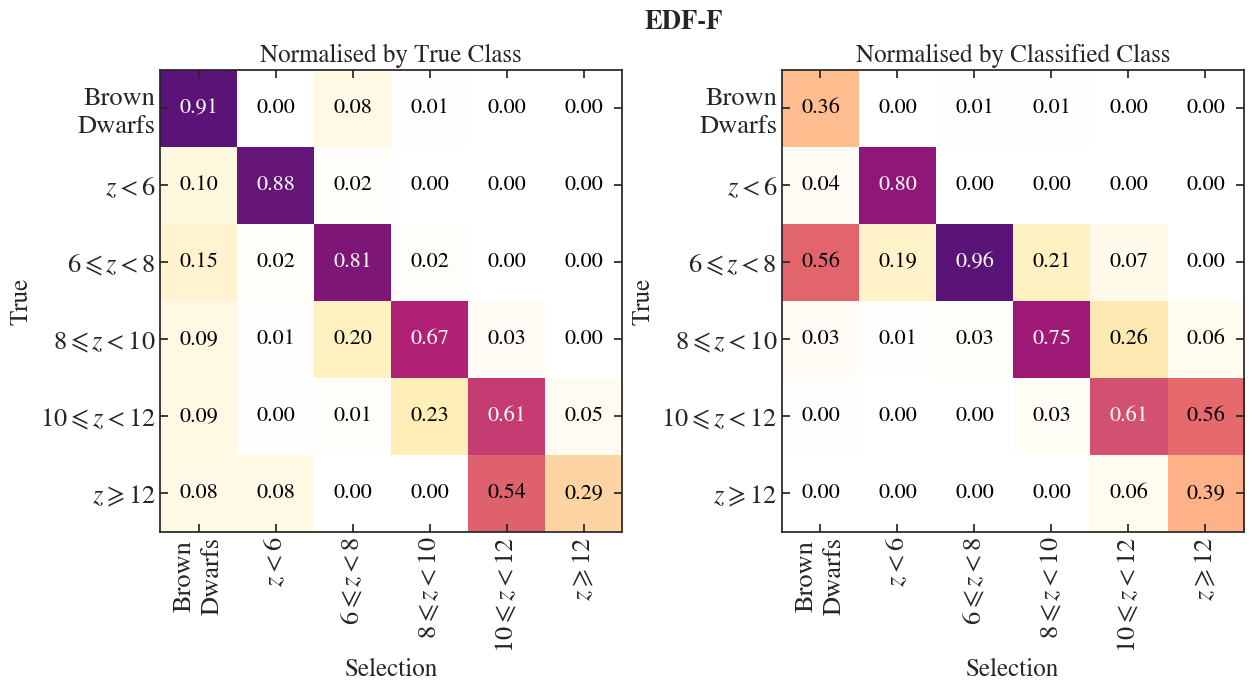}
         \label{fig:recoved_sel_frac_cdfs}
     \end{subfigure}
     \hfill
     \begin{subfigure}{\textwidth}
         \centering
         \includegraphics[width=0.78\textwidth]{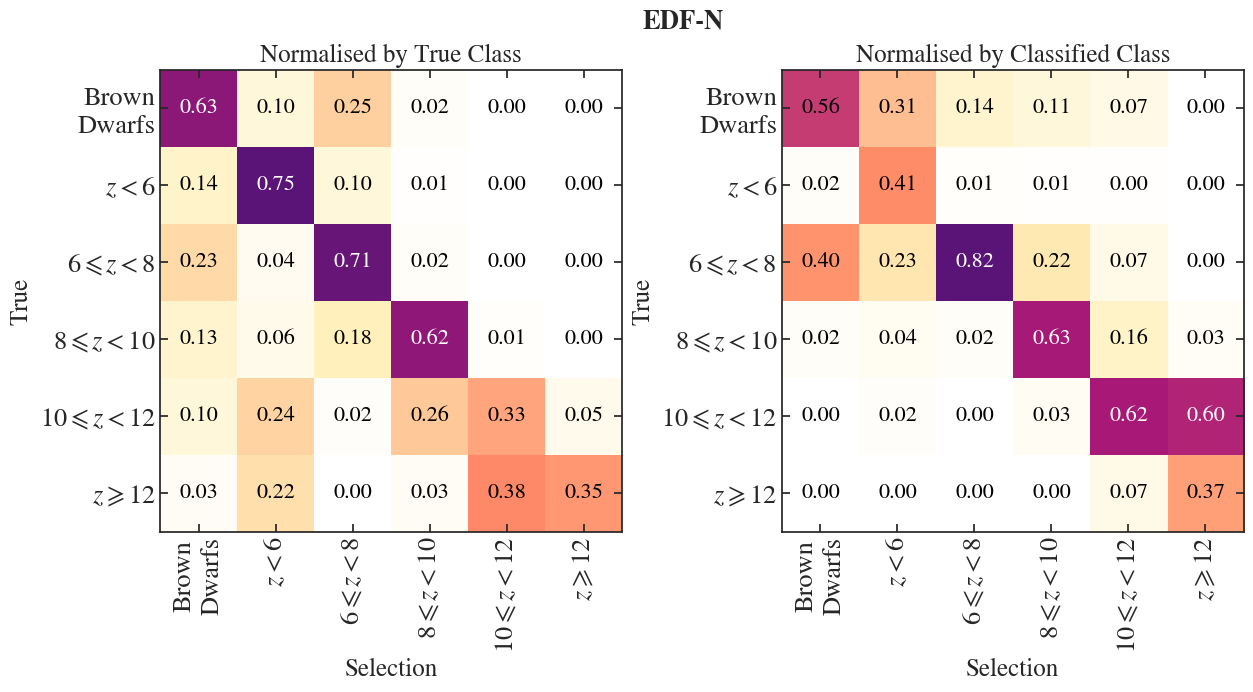}
         \label{fig:recoved_sel_frac_nep}
     \end{subfigure}
        \caption{Confusion matrices, for each EDF, presenting the performance of our classification (Sect. \ref{sc:EF_Paper_Method}). The left matrices are normalised by true class, and the right matrices are normalised by selected class. Diagonals present the completeness (\textit{left}) and purity (\textit{right}). }
        \label{fig:recoved_sel_frac_allfields}
\end{figure*}

\section{\label{sc:EF_Paper_Results} Results}

\begin{figure*}[t]
    \centering
    \includegraphics[width=1\linewidth]{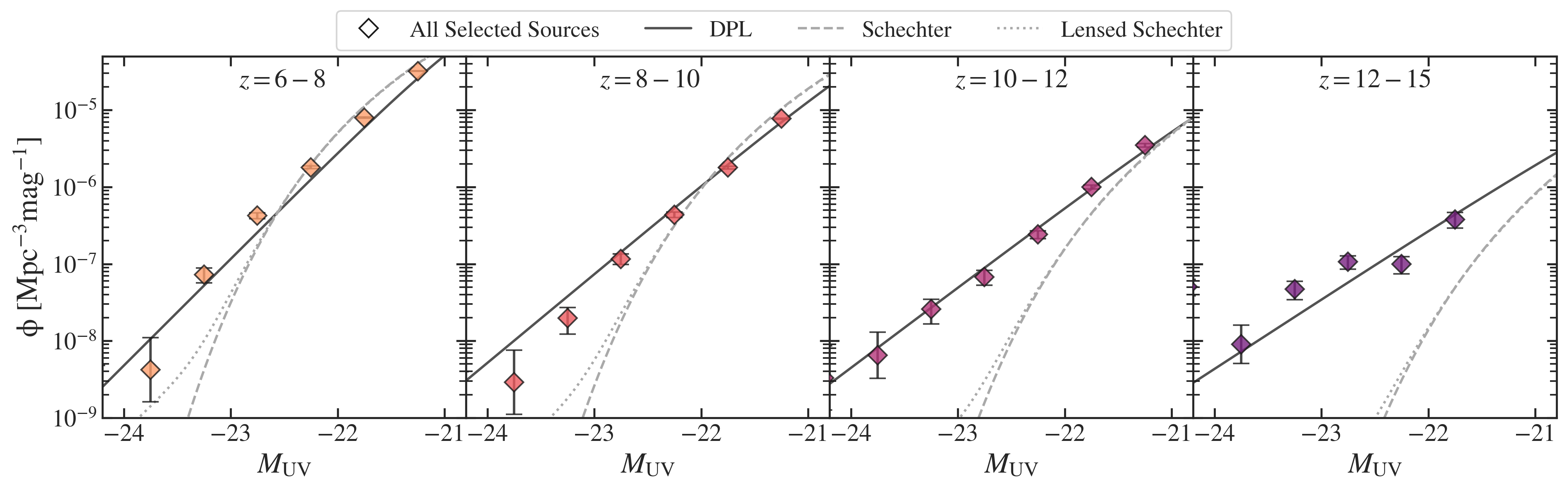}
    \caption{Combined EDF UV LF of the samples classified in four redshift bins: $6\le z<8$, $8\le z<10$, $10\le z<12$, and $12\le z<15$. Here, $M_{\rm UV}$ is calculated in \JE band for $z<10$ and \HE band for $z>10$, using Eq. (\ref{equ:MabsCal}). Number densities for all sources classified in each $z$-bin are shown as diamonds, with error bars assuming Poisson statistics. 
    The solid black, grey dashed, and grey dotted lines show the input catalogue DPL model, Schechter model from \citet{whitler25_jades_uvlf} (JWST-based measurements) and a lensed Schechter model (using \citealt{Mason2015_UVLFLensing}), calculated at the median $z$ of the $z$-bin. With these measurements, we can distinguish between DPL and Schechter models, highlighting the importance of \Euclid in constraining the bright end of the UV LF.}
    \label{fig:edf_uvlf}
\end{figure*}

\subsection{\label{sc:EF_Paper_Results_RecoveryPurity}Performance of selection criteria} 

This paper tests various selection methods on three synthetic catalogues to determine the optimal criteria for selecting $z\ge 6$ galaxies in the EDFs, while minimising contamination. The optimal criteria, described in Sect. \ref{sc:EF_Paper_Method}, combines SED fitting with S/N cuts in the \IE band. Using the best-fit photometric redshift from the SED fitting, we classify sources in our catalogue into classes: MLTs, $z<6$, $6\le z<8$, $8\le z<10$, $10\le z<12$, and $z\ge 12$ galaxies. The results are presented in the confusion matrices in Fig. \ref{fig:recoved_sel_frac_allfields} for each EDF. The matrices shown on the left have been normalised over rows (``true'' input values), and the matrices on the right have been normalised over columns (``classified'' sample). Therefore, the diagonals in the left matrices present the completeness of each source type, while in the right matrices, the diagonals present the purity of each class. The completeness, $R$, of a class, $i$, is described by

\begin{equation}
    R_i = \frac{N(T=i, C=i)}{\sum_j N(T=i, C=j)}.
    \label{equ:completeness}
\end{equation}
Here, $i$ and $j$ are the labels/classes: MLTs, $z<6$, $6\le z<8$, $8\le z<10$, $10\le z<12$, and $z\ge 12$ sources. $N(T,C)$ is the number of objects of true class $T$ and predicted class $C$. Therefore, the numerator, $N(T=i, C=i)$, is the total number of sources with truth label $i$, classified in class $i$. The denominator, $\sum_j N(T=i, C=j)$, is the total number of sources of class $i$ in all classes.

The purity of a class, $P_i$, is described as

\begin{equation}
    P_i = \frac{N(T=i, C=i)}{\sum_j N(T=j, C=i)}.
    \label{equ:purity}
\end{equation}
Here, the denominator is the total number of sources classified in the class, $i$. 

Before discussing our results, we will first explain the matrices in Fig. \ref{fig:recoved_sel_frac_allfields} through examples. The top left matrix of Fig. \ref{fig:recoved_sel_frac_allfields} shows that our criteria can recover $80\%$ of the true $6\le z<8$ sources (third row and third column from the top left corner). Looking along the row of $6\le z<8$ sources (third row) shows that the remaining $20\%$ have been misclassified as other types of sources. For example, $16\%$ of true $6\le z<8$ sources were misclassified as MLTs (third row, first column). 
In the top right matrix of Fig. \ref{fig:recoved_sel_frac_allfields}, the diagonal shows that $96\%$ of objects classified as $6\le z<8$ sources are true $6\le z<8$ sources. By looking along this column, we can also see that $2\%$ of this class are MLTs. Next, we will discuss the main results that these matrices describe.

For EDF-F and EDF-S, the completeness decreases with increasing redshift; the completeness is lowest in true $z>10$ sources. This is shown along the diagonals of the left matrices for EDF-S and EDF-F in Fig. \ref{fig:recoved_sel_frac_allfields} (top and middle). Our results show that, for both fields, $62\%$ of the true $z\ge 12$ sources were classified as $10\le z<12$ sources (sixth row) in EDF-S, and $54\%$ were missclassified in EDF-F. This is due to $z>10$ sources being faint, such that their errors in {\it Spitzer}/IRAC and ground-based optical photometry are larger, leading to a wider photometric posterior in the SED fitting process. 

As shown in the diagonals of the right matrices of Fig. \ref{fig:recoved_sel_frac_allfields} (top and middle), the purity in EDF-S and EDF-F decreases with increasing redshift for classes: $6\le z<8$, $8\le z<10$, $10 \le z<12$, and $z>12$. In these classes, the contamination is from high-$z$ sources outside of the chosen redshift bin, rather than $z<6$ interlopers or MLTs. For example, the sample classified as $z>12$ sources has a significantly large contamination of $10\le z<12$ sources ($56\%$ in EDF-S and EDF-F). This ``contamination'' from high-$z$ sources affects the $z>12$ UV LF due to the additional correction in the UV magnitudes (see Sect. \ref{sc:EF_Paper_Results_UVLF}).

For EDF-N (Fig. \ref{fig:recoved_sel_frac_allfields}, bottom), we find similar trends to EDF-S and EDF-F, in completeness and purity. However, these values are generally lower, especially for $z>10$ sources. For example, as shown in the diagonal of the bottom left matrix, the completeness of $10\le z<12$ sources in EDF-N is $33\%$, while for EDF-F, it is $61\%$. This is due to the shallower optical data and the lack of {\it Spitzer}/IRAC photometry in half of the EDF-N field, resulting in broader photometric posterior distributions and confusion between the Balmer and Lyman breaks. This is shown along the final row of the bottom left matrix in Fig. \ref{fig:recoved_sel_frac_allfields}, where $22\%$ of true $z>12$ sources are classified as $z<6$ sources, and $38\%$ of true $z>12$ sources are misclassified as $10\le z<12$ sources.

EDF-N also has higher contamination levels of MLTs in each class, compared to the other fields. This is shown in the top row of the bottom right matrix in Fig. \ref{fig:recoved_sel_frac_allfields}. This is likely a result of the larger number densities of MLTs in EDF-N (see Fig. \ref{fig:browndwarf_nomdens_per-field}) as well as the lower area coverage of {\it Spitzer}/IRAC photometry in this field, which is important in SED fitting to disentangle MLT and galaxy best-fit templates.

The completeness and purity of our sample are affected by magnitude. For all EDFs, the completeness and purity rates are almost perfect when considering sources with $M_{\rm H} < 25$. For fainter sources, the rates are similar to those shown in Fig. \ref{fig:recoved_sel_frac_allfields}; fainter sources are difficult to recover due to larger photometric errors, which widen the photo-$z$ posterior distribution.

\subsection{\label{sc:EF_Paper_Results_UVLF}Luminosity function}

Using our selection, we now estimate how well the UV LF in the EDFs can be recovered.
The LF of the final sample are measured in 4 redshift bins: $z=6$--8, $z=8$--10, $z=10$--12, $z=12$--14 using the \texttt{eaz} best-fit photometric redshifts. In absolute magnitude bins of $\Delta M = 0.5$, the LF is derived using 
\begin{equation}
    \phi(M, z) = \frac{N(M, z)}{V_{\mathrm{eff}}(M, z) \; \Delta M},
\end{equation}
where $N$ and $V_{\mathrm{eff}}(M, z)$ are the number of sources selected and the effective volume, in a given redshift ($z$) and absolute magnitude, $M$, bin. 

The effective volume is calculated in each redshift bin by
\begin{equation}
    V_{\mathrm{eff}}(M) = \int^{z_{\rm max}}_{z_{\rm min}} S(M, z)\, A \, \frac{\mathrm{d}V(z)}{\mathrm{d}z}\, \mathrm{d}z.
\end{equation}
Here, $S(M, z)$ is the selection function, $A$ is the area in angular units, and $\frac{\mathrm{d}V(z)}{\mathrm{d}z}$ is the differential co-moving volume at a given $z$. The selection function is the fraction of (known) high-$z$ sources selected, divided by the number of sources expected in that given $z$ and $M$ bin. To derive this, we create a photometric catalogue with uniform number counts of LBGs over the redshift range of $5\le z < 15$, using the same templates as discussed in Sect. \ref{sc:EF_Paper_Data_highz}. We run \texttt{eaz} and apply the same selection criteria to this catalogue, as described in Sect. \ref{sc:EF_Paper_Method}. We bin the full catalogue and final selected sample in both (photometric) redshift ($\Delta z = 0.1$) and absolute magnitude ($\Delta M_{\rm UV} = 0.1$), where $M_{\rm UV}$ is calculated using the photometric redshifts. The selection function is obtained by dividing the final binned and input binned data by each other. 

The absolute magnitudes of our sample are calculated as
\begin{equation}\label{equ:MabsCal}
    M_{\rm UV} = m - \mathrm{DM}(z) + 2.5\, \log_{10}(1+z) + \Delta M_{\rm loss}(z), 
\end{equation}
where $m$ is the apparent magnitude in the band encompassing $1500\, \AA$. We use the \JE band for $z<10$ and \HE for $z>10$. Due to the break falling in the \HE-band at $z>10$, we include $\Delta M_{\rm loss}$ to account for the flux lost in the \HE band. Finally, $\mathrm{DM}(z)$ is the distance modulus at the source's (photometric) redshift, $z$, which is calculated using \texttt{astropy.cosmology}.

The number densities of our final selection for all EDFs are presented in Fig. \ref{fig:edf_uvlf}. For the $z<12$ LFs, the measurements are within 1--2$\sigma$ of the input DPL, shown as the solid black line. Deviations from the input could potentially be explained by contamination. 
However, when removing the contaminants, we find that our values are within $2\sigma$ of the high-$z$ only counts; therefore, the final number counts are not significantly affected by contamination of low-z sources and MLTs. 
For the $z=12$--15 LF, the diamonds and squares deviate from the DPL model at $M_{\rm UV} < -23$ due to sources from $z\sim 10$ scattering into this bin, for which the luminosity is then significantly overestimated (see also Sect. \ref{sc:EF_Paper_Results_RecoveryPurity}).  

\subsection{Number of expected and selected sources}

\begin{figure}
    \centering
    \includegraphics[width=0.9\linewidth]{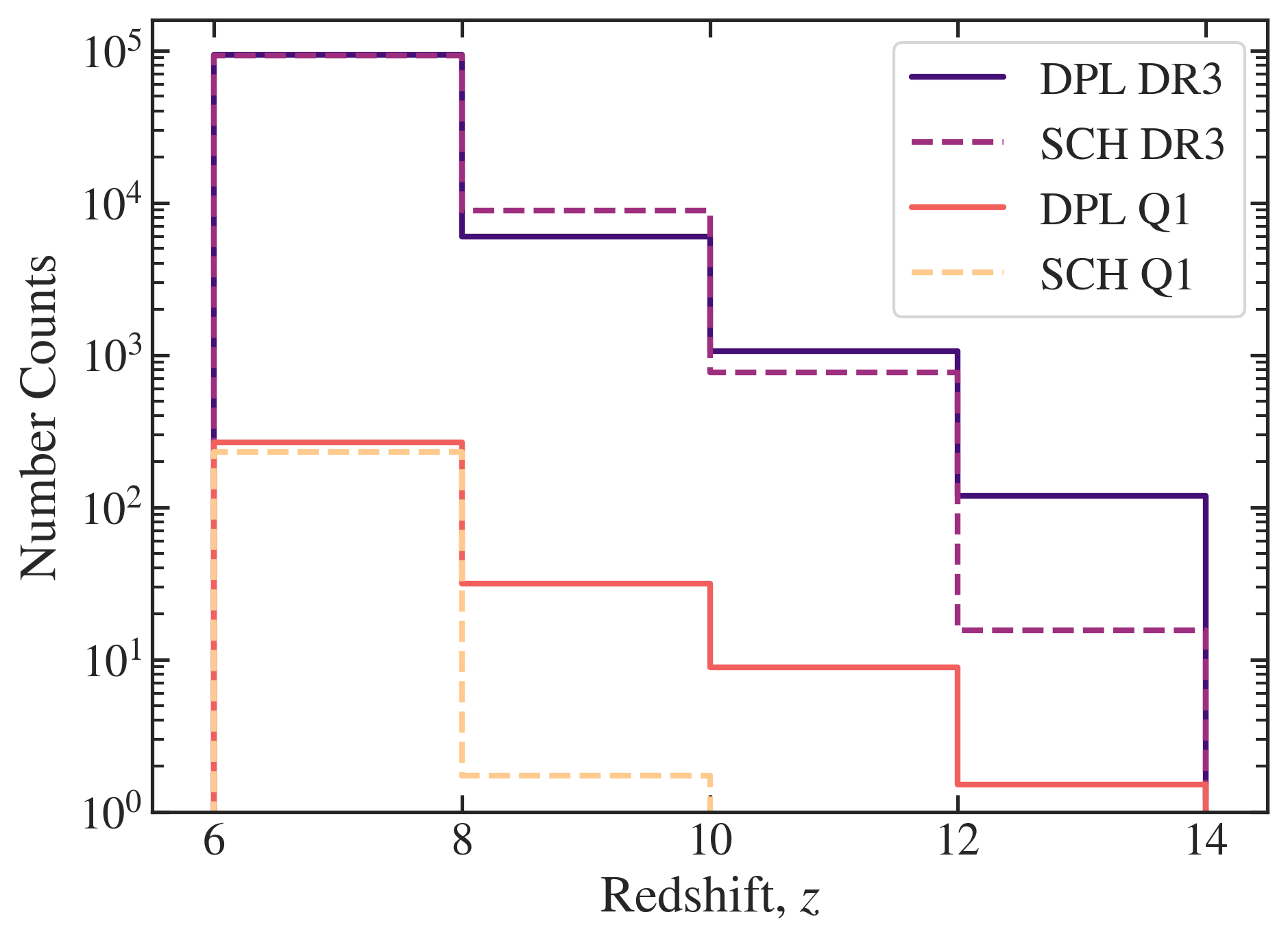}
    \caption{Expected number counts of $z>6$ LBGs per $\Delta z=1$ for DR3 (26.5 AB depth) and Q1 (24.5 AB depth) estimated from the DPL (solid) and Schechter (dashed) LFs. With both \Euclid datasets, we will be able to distinguish between these models and constrain the bright end of the UV LF.}
    \label{fig:expnumcounts_hist}
\end{figure}

The number of sources that are expected to be observed in the final \Euclid data release and the first Quick release (Q1; \citealt{Q1cite, Q1_Release}) are shown in Fig. \ref{fig:expnumcounts_hist}, assuming the DPL LF from \citet{bowler2020_uvlf_ultravista} at $z<8$ and \citet{donnan24_jwst_uvlf} at $z\ge8$ as well as the Schechter LF from \citet{whitler25_jades_uvlf}. 

At the depth of the Q1 dataset, we do not expect to detect any $z>8$ sources if the UV LF follows the Schechter model. However, if the UV LF follows the DPL then this redshift limit increases to $z\sim 12$. Therefore, with the first \Euclid data, we will be able to distinguish between the Schechter or DPL models (see Sect. \ref{sc:EF_Paper_Discussion:SCH_DPL} for more details).

For the final \Euclid data release (DR3), we expect similar number counts of $z<11$ sources with the DPL and Schechter UV LF models. However, at $z>11$, the number counts deviate between the models, and we expect more than twice the number of sources with the DPL model than with the Schechter model. This highlights the importance of the EDFs in constraining the bright end of the UV LF at early epochs. The expected values of $z>6$ LBGs are presented in Table \ref{tab:dr3_highz_numcounts} (in columns labelled ``DPL'' and ``Schechter'') and in the Appendix for Q1 estimates (see Table \ref{tbl:Q1_expectednumbercounts}).

However, these expected number counts do not account for selection and observational effects, which we discuss further in Sect. \ref{sc:EF_Paper_Discussion_Validity}.
The column labelled ``Selected (DPL)'' in Table \ref{tab:dr3_highz_numcounts} presents the number counts of LBGs selected using the method described in Sect. \ref{sc:EF_Paper_Method}. For the majority of the redshift bins, we select $>75\%$ of the expected sources.  With these selected values, we will be able to distinguish between the DPL and Schechter models. However, for the $8\le z < 10$ bin, we obtain more sources than expected for the DPL model. This is due to contamination from $6\le z<8$ sources as well as some MLTs (from the EDF-N field). 

\begin{table}[]
    \centering
    \caption{The expected number of $z>6$ sources (with poison errors) for a 53 deg$^2$ survey, with depth down to 26.5 AB (e.g. DR3), estimated from the DPL and Schechter UV LF models The ``Selected'' column presents the total number of sources (high-$z$ and contaminants) classified in each redshift bin using the optimal method, described in Sect. \ref{sc:EF_Paper_Method}. For Q1 estimates, see Table \ref{tbl:Q1_expectednumbercounts} in the Appendix.}
\setlength{\tabcolsep}{0.5em} 
{\renewcommand{\arraystretch}{1.3}
    \begin{tabular}{c|c|c|c}
       Redshift &  DPL & SCH & Selected (DPL)\\ \hline \hline
        $6\le z < 8$ & $93\,893 \pm 306$ & $92\, 435 \pm 304$ & 70\,445$ \pm {265}$\\
        $8\le z < 10$ & $6016 \pm 78$ & $8905 \pm 94$ & 8370$ \pm {91}$ \\
        $10\le z < 12$ & $1066 \pm 33$ & $770 \pm 28$ & 971$ \pm {31}$\\
        $12\le z < 14$ & $119 \pm 11$ & $16 \pm 4$ & 113$ \pm {11}$\\
    \end{tabular}}
    \label{tab:dr3_highz_numcounts}
\end{table}

\section{\label{sc:EF_Paper_Discussion} Discussion}

\subsection{\label{sc:EF_Paper_Discussion_Validity}Validity of high-$z$ selection criteria}
In this section, we discuss the validity of our selection criteria and their use for the real \Euclid data releases. This work has measured the contamination from low-$z$ interlopers and MLTs on the final sample, but does not consider contaminants such as extremely red low-$z$, quiescent sources and AGN/quasars (see Sect. \ref{sc:EF_Paper_Discussion-AGN} for details on AGN contamination). These sources have been studied extensively in \citet{vanMierlo-EP21}, who finds a larger contamination fraction ($\sim 13\%$ in total) after including them.

In real datasets, there are further contamination/issues that we could not include in our synthetic catalogues. This includes source deblending (e.g. in \Spitzer IRAC), removal of artefacts or persistence. With the ERO data and Q1 release, artefacts such as ghosts and persistence have been identified in these first sets of VIS and NISP images (e.g. \citealt{Weaver_EROLensVISDropouts}). There are also difficulties in obtaining {\it Spitzer}/IRAC photometry for high-$z$ sources if there are close neighbours, especially bright ones. Therefore, our synthetic catalogues present a ``perfect'' dataset, and the errors on our number densities may be underestimated due to these missing issues. Extra selection criteria will be needed to remove these artefacts, for example, visual inspection. This is discussed further in Sect. \ref{sc:EF_Paper_Discussion_Q1Examples} as well as works by Weaver et al. (in prep.).

\subsection{\label{sc:EF_Paper_Discussion:SCH_DPL}Distinguishing between Schechter and DPL rest-UV luminosity function}

Prior to the launch of \Euclid, the current volumes probed at high redshift, e.g. by JWST or HST, are too small to constrain the bright end of the UV LF (where areas are $\rm < 1 deg^2$). The bright end of the $z>4$ UV LF has been measured using ground-based instruments, e.g. Subaru/HSC and UltraVISTA (see, for example \citealt{Stefanon2019_BrightCosmosz8UVLF, bowler2020_uvlf_ultravista, harikane_goldrish_uvlf, Adams2023_AGN_Gal_UVLF}). However, ground-based telescopes are affected by atmospheric effects, weather and have long survey times. Subaru/HSC is limited in wavelength and can probe up to $z=7$. UltraVista can detect galaxies up to $z=13$ with the {\it K}-band, but the volumes probed at high-$z$ are small. This reduces the number of bright galaxies that can be detected, leading to large errors in the number counts of bright high-$z$ sources. 

With \Euclid we are entering an era where we can obtain wide and deep imaging in NIR wavelengths, which will further push the redshift boundary to $z=14$, helping us to constrain the bright end of the high-$z$ UV LF. In Fig. \ref{fig:edf_uvlf}, we present the number densities of our final selection in four redshift bins. In each panel, we have also included the Schechter model from the JWST-based study \cite{whitler25_jades_uvlf} for comparison. With our criteria, we can recover the DPL within 1--$2\sigma$. The error bars on our number densities are also small enough to be able to distinguish between the Schechter function at $ M < -22$ AB, showing that the \Euclid mission will help us to constrain the UV LF up to $z\sim 14$. However, at these bright magnitudes, AGN contribution may come into play in the real \Euclid data releases. We discuss the effects of AGN in the following section (Sect. \ref{sc:EF_Paper_Discussion-AGN}).  In Fig. \ref{fig:edf_uvlf} we have also included the model from \citet{Mason2015_UVLFLensing}, which includes the effects of lensing by foreground galaxies. We find that this lensing effect does not increase the number densities of sources within the volume probed by the EDFs. These effects come into play in areas as large as the \Euclid wide survey.

\begin{figure*} 
     \centering
     \begin{subfigure}[b]{1\textwidth}
         \centering
         \includegraphics[width=1\textwidth]{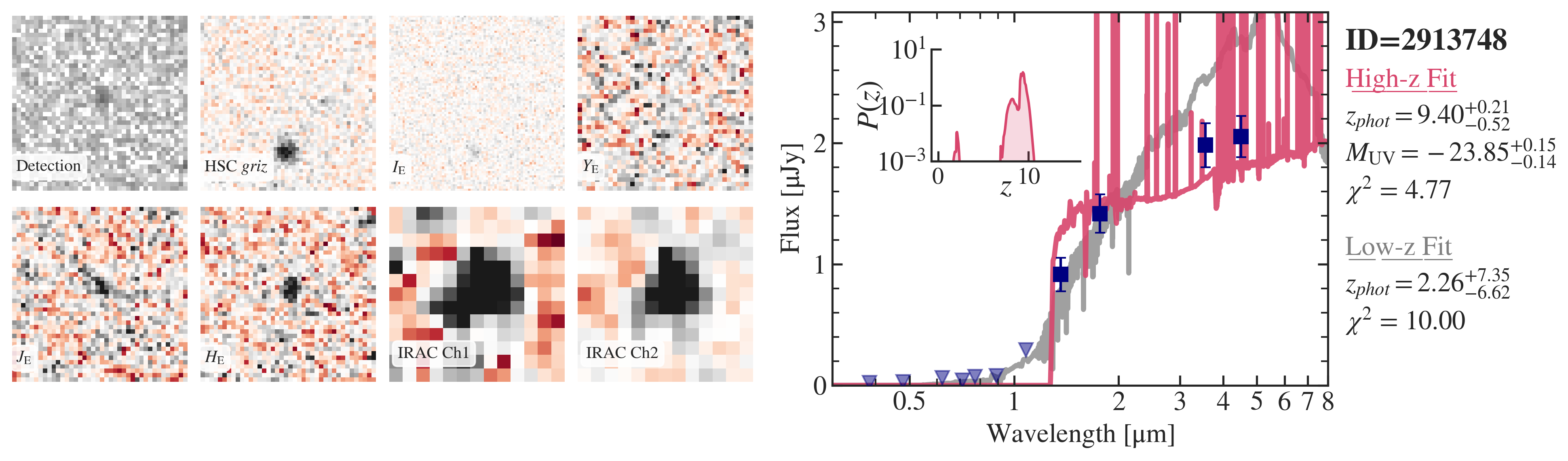}
     \end{subfigure}
     \hfill
     \begin{subfigure}[b]{1\textwidth}
         \centering
         \includegraphics[width=1\textwidth]{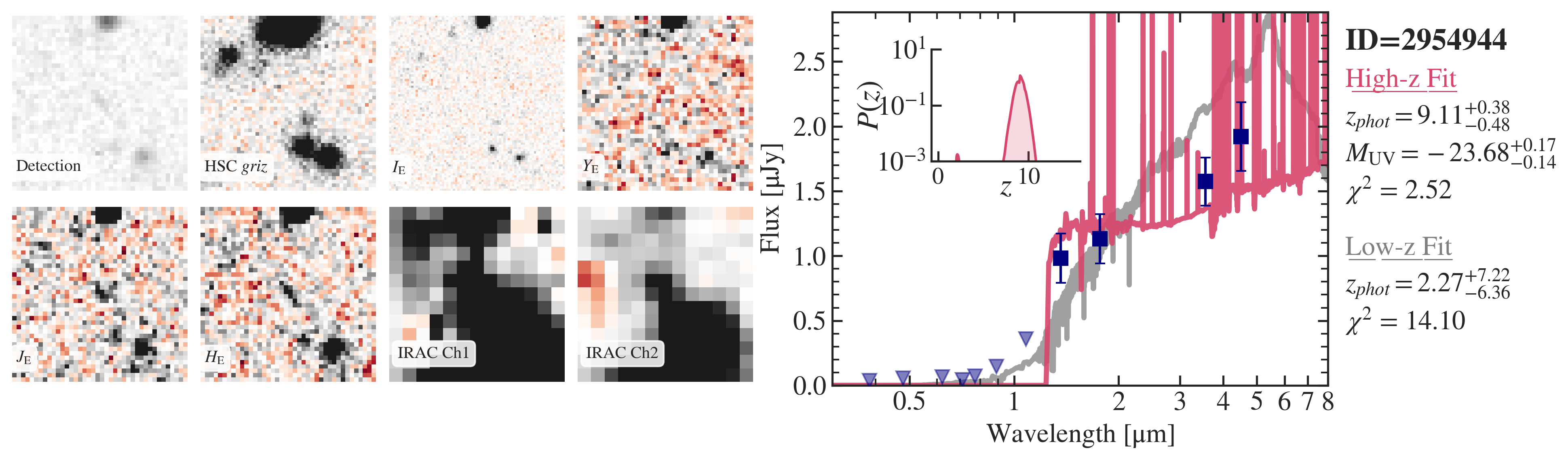}
     \end{subfigure}
    
        \caption{Two ultra-bright $z>8$ candidates selected from the EDF-N Q1 DAWN catalogue (\citealt{Q1_Release}, Weaver et al. in prep.). \textit{Left}: $8''$ size cutouts centred on detected source in multiple bands: CFHT-$u$, HSC stacked $griz$, \IE (MER tile), \YE\JE\HE detection, coloured RGB, \YE, \JE, \HE, IRAC Ch1, and IRAC Ch2. Middle: Best-fit SED templates from \texttt{eaz}, when fit over the redshift range $0<z<15$ (`full'; red) and $0<z<6.5$ (`low-$z$'; grey). Non-detections are shown as triangles at their 2$\sigma$ upper limits. \textit{Right}: Logged probability distribution of the best-fit photometric redshift from \texttt{eaz}, for the SED fitting over $0<z<15$. }
        \label{fig:q1_examps}
\end{figure*}

\subsection{\label{sc:EF_Paper_Discussion-AGN}Contamination from AGN and Quasars}

The work discussed in this paper focuses on the selection of LBGs at $z>6$, but does not consider the contribution from faint-AGN or the contamination from quasars. Studies on the AGN and galaxy UV LF have shown that AGN start to dominate the light from sources at $M_{\rm UV} < -23$ (e.g. \citealt{Ono18_uv_lf,harikane_goldrish_uvlf, Finkelstein2022_AGNGalaxyUVLF}), which may be contributing to the overabundance of bright high-$z$ sources found by recent JWST studies. Although work from e.g. \citet{Finkelstein2022_AGNGalaxyUVLF} found that the number densities of bright candidates still follow a DPL model after accounting for AGN contribution, suggesting other mechanisms may lead to the overabundance. AGN contamination is believed to only affect high-z samples at $z<7$, as recent work from e.g. \citet{Pratika2025_JWSTALMA_AGN}, found that AGN number densities become important at $z<7$.

For quasars, prelaunch \Euclid works from \citet{selwood_agn-nomdens} and \citet{barnett17_quarars_z79} estimate significant number counts of $z<7$ and $z>7$ quasars, respectively, to be selected in the Euclid Wide Survey. However, at higher redshifts, the number densities of quasars fall, and they become rarer to select (e.g. \citealt{Schindler23_QuasarLF}). Therefore, quasars may not dominate the contamination in high-$z$ galaxy samples.

Removing contamination from AGN and quasars is difficult, especially at high-$z$, where the UV SED looks similar to high-$z$ galaxies. One way to remove these sources is by using morphology, as bright AGN and quasars have point-source morphology, but this will become more difficult for fainter sources.
The use of follow-up spectroscopic observations, e.g. ground-based or JWST, of these \Euclid sources will help to determine AGN contributions using emission line width. 

\subsection{Alternative criteria}

With our synthetic catalogue, we tested various methods to recover the optimal selection criteria. Here, we will summarise the results of these tests. For further details, we direct the reader to Appendix  \ref{sc:EF_Paper_Discussion_OtherCrietria}. 

Combining \Euclid with ground-based and {\it Spitzer} photometry is important to increase the recovery rates of $z>6$ sources. {\it Spitzer}/IRAC is important for recovering $z>10$ sources, but ground-based data is further required to constrain the Lyman break and reduce confusion between the Balmer break of low-$z$ template solutions. 

Finally, further cuts using the ground-based photometry are not required to obtain a pure sample of high-$z$ candidates. SED fitting is more important for the removal of MLTs, and further S/N cuts using all optical bands reduce the recovery of high-$z$ sources more than MLTs.

\section{\label{sc:EF_Paper_Discussion_Q1Examples}High-$z$ candidates from Q1 datasets}

\begin{figure}
    \centering
    \includegraphics[width=0.9\linewidth]{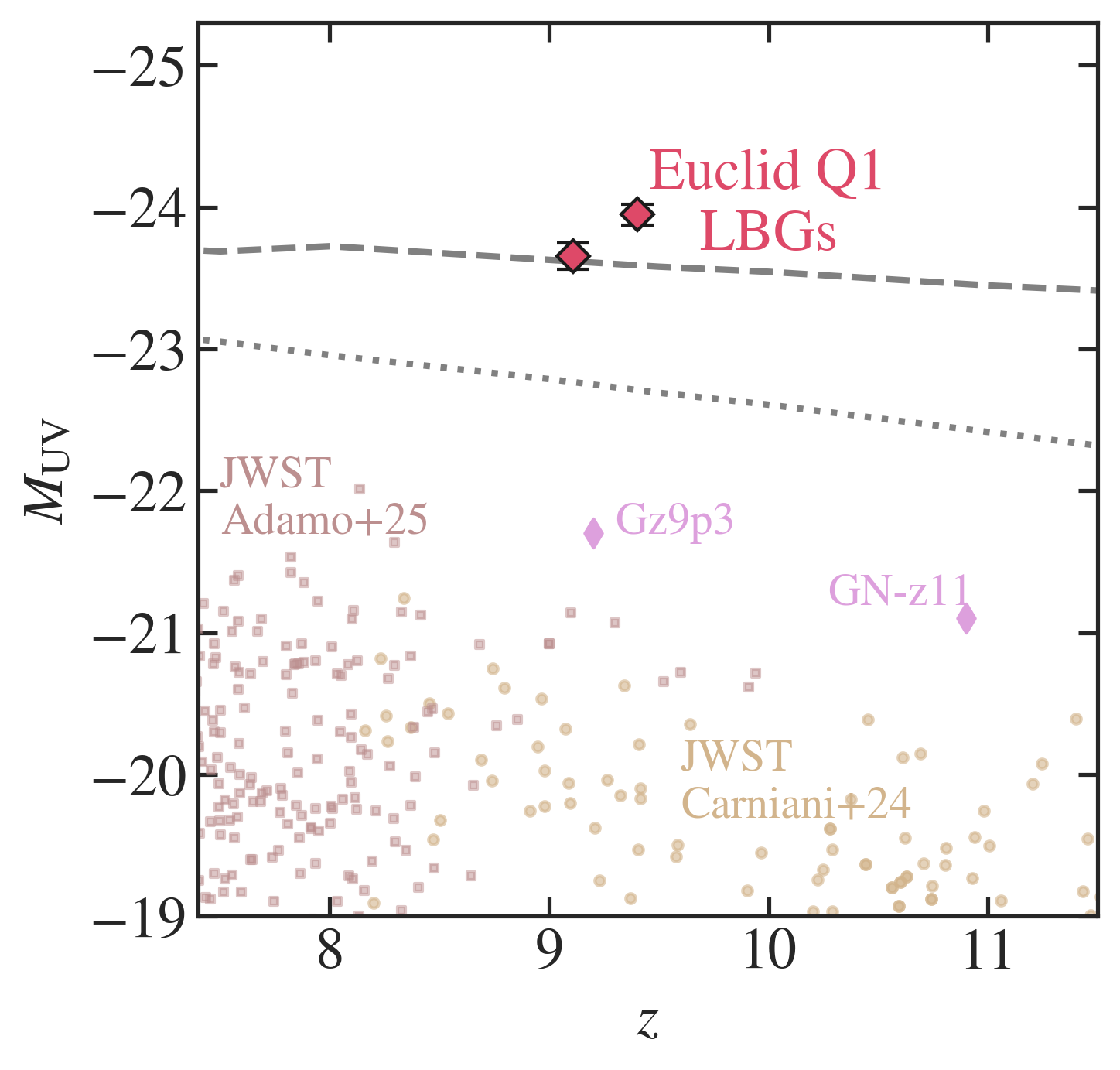}
    \caption{Absolute magnitude ($M_H$ at $z_{\rm phot}$ and photometric redshift from the best-fit SED templates from eazy, for a two $z>9$ sources selected in the EDF-N Q1 dataset. These sources were selected using conservative criteria and thus are not representative of the whole $z>8$ population. Grey dashed (dotted) lines show the magnitude at which we expect one galaxy per redshift bin for the DPL (Schechter) LF models used in this paper in the 10 deg$^2$ volume that the \Euclid Q1 data probes at the moment in the EDF-N. Sources from the \citealt{Adamo24_review} and \citealt{Carniani24} surveys are shown as brown and tan points. Gz9p3 and GN-z11 are also plotted in grey for reference.}
    \label{fig:q1_examps_Muv_z}
\end{figure}

In this section, we present two $z>8$ ultra-bright candidates from the Q1 dataset, selected using a similar method to the one in this paper but adapted for purity. These candidates were selected from the Cosmic Dawn \Euclid Farmer catalogue (see \citealt{EP-Zalesky, EP-McPartland} and  Weaver et al. in prep for details)

To select a sample of $z>8$ sources from the Q1 dataset, a similar method to the one described in Sect. \ref{sc:EF_Paper_Method} is used. However, to select a pure and higher redshift sample from these early data release, we adapt this method to include S/N cuts of $\mathrm{S/N}(\YE) < 2$, $\mathrm{S/N}(\JE) > 5$, and $\mathrm{S/N}(\HE) >5$ as well as $\mathrm{S/N}<2$ cuts in all HSC bands. We also adopt the recommended masking criteria of this catalogue. The photometric redshifts and other physical properties of the Q1 sources were determined with the SED fitting software \texttt{LePhare} (\citealt{lephare}). Therefore, the sample is selected initially using the \texttt{LePhare} $z_{\rm phot}$, $\chi^2_{\rm stellar}$, and $\chi^2_{\rm galaxy}$. This criterion obtains over $100$ sources, which are visually inspected to select a sample of 31 candidates. Most of these sources were artefacts, e.g. persistence and ghosts in the Q1 NISP images. SED fitting from \texttt{eazy} is run on these 31 candidates to remove any further contaminants. To obtain a pure sample at $z>8$, we select sources that are selected as $z>8$ sources in both \texttt{eazy} and \texttt{LePhare} SED fitting and have 84\% probability of being at $z>8$. After applying this criterion, we obtain a sample of 9 plausible candidates from EDF-N, which is consistent within $1\sigma$ of the expected number counts (see Table \ref{tbl:Q1_expectednumbercounts}) assuming Poisson errors. This method aims to select a highly pure sample rather than a complete sample; therefore, this is significantly more conservative than the method discussed in Sect. \ref{sc:EF_Paper_Method}.

Cutouts and SED fittings of the two most secure ultra-bright $z>8$ candidates from the 10~$\mathrm{deg}^2$ of EDF-N are presented in Fig. \ref{fig:q1_examps} as examples. Candidate 449\,994, in Fig. \ref{fig:q1_examps}, has a large excess in IRAC Ch2. This excess may be a result of contamination from neighbours in the IRAC Ch2 image, but Farmer has been optimised to accurately model fit these images (\citealt{FarmerSoftware, Weaver23_Farmer}). Thus, if such an excess is accurate, this candidate either has extreme optical emission lines or a large Balmer break. 
As shown in Fig. \ref{fig:q1_examps_Muv_z}, these candidates are brighter than sources from other literature, such as samples presented in (\citealt{Adamo24_review}) and (\citealt{Carniani24}). In comparison to JWST sources Gz9p3 (\citealt{Gz9p3_Nature2024}) and GN-z11 (\citealt{Oesch2016_GNz11_Spec, Bunker2023_GNz11_JWST}), our sources are over two magnitudes brighter, showing the capabilities \Euclid has in finding the brightest and rarest candidates in the early Universe. 
The dashed and dotted grey lines in Fig. \ref{fig:q1_examps_Muv_z} present the magnitude expected for one galaxy per redshift bin of $\Delta z=1$, in a survey area of 10\,deg$^2$,  for the DPL and Schechter models considered in this work, respectively. 
Our candidates lie within 2$\sigma$ of the DPL model, consistent with the overabundance of bright high-$z$ sources found by JWST. 
Spectroscopic follow-up is needed to confirm their redshifts. This is possible with, e.g. MOSFIRE, to detect the Ly$\alpha$ emission lines, or follow up with JWST NIRSpec to obtain high-resolution rest-optical lines, which can be used to address any possible AGN contribution. If the photometric redshifts are accurate, we will find tens of thousands of these $z>6$ sources by the end of the \Euclid 6-year mission.

\section{\label{sc:EF_Paper_Conclusion} Summary}

The bright end of the rest-UV LF provides unique constraints for understanding the evolution of the brightest and rarest galaxies in the Universe. Many studies have attempted to constrain the bright end at $z>6$ using both ground- and space-based missions. Current NIR space-based missions are limited by area, and ground-based telescopes are limited by depth, long survey times, and weather effects. Now with \Euclid, we will obtain data over 53\,deg$^2$, down to $26.5\, \rm AB$, allowing us to fully constrain the bright end at $z>6$. However, it is difficult to select these high-$z$ sources due to contamination of $z<6$ interlopers and MLTs, which can mimic the colours of high-$z$ LBGs in broadband filters.

In this work, we create three mock catalogues that simulate the data of the EDFs at the end of the 6-year mission. These catalogues were used to test various criteria to select $z>6$ sources with the highest completeness and purity. We find the following results:
\begin{enumerate}
    \item The most optimal method for selecting $z>6$ LBGs uses a combination of S/N cuts with SED fitting. The majority of low-$z$ interlopers that we consider in our synthetic catalogue are removed with the \IE S/N cut, leaving a contamination rate of less than $10\%$ in EDF-N and $2\%$ in EDF-S and EDF-F. 
    
    \item SED fitting of stellar templates is important to remove $90\%$ of MLTs from the final sample. Further cuts in the ground-based optical bands can improve the purity of the final sample, but the completeness of $z>6$ sources drops by $\sim 10\%$, due to up-scattering in the optical bands and the residual Lyman alpha forest signal in some $z<6.5$ LBGs.
    
    \item To obtain the highest completeness and purity of $z>6$ samples, all auxiliary data should be included in the SED fitting process.  Optical and MIR coverage is crucial for distinguishing the Lyman and Balmer breaks of high-$z$ sources. MIR is also important to include in selecting $z>10$ sources and removing MLTs.
    
    \item Based on current DPL LF models, we expect to find more than $ 100\,000 $ and $100$ source at $z=6$--12 and $z>12$ in the final \Euclid data release. Assuming the recent JWST-based Schechter models, we expect similar number counts of $z=6$--12 galaxies to the DPL model, but fewer than $16$ sources at $z>12$. This highlights the importance of the \Euclid mission in constraining the bright end of the UV LF in the very early Universe. 
    
    \item Using our optimal selection, we obtain more than $75\%$ completeness at $z>6$. Although at $8<z<10$, we select more than the expected value, due to contamination of other high-$z$ sources (rather than low-$z$ interlopers and MLTs).
    
    \item We present two ultra-bright $z>8$ candidates identified in 10 deg$^2$ of the \Euclid Q1 dataset.
    These sources are $\sim 1$--2 magnitudes brighter than sources from BORG and REBELS surveys, as well as other bright high-$z$ sources such as GNz11 and Gz9p3. If the photometric redshifts of these sources are accurate, we expect tens of thousands more bright sources by the end of the \Euclid mission.
    
\end{enumerate}
\begin{acknowledgements}
\AckEC  
\AckQone
This work is produced with funding from the Danish National Research Foundation under grant DNRF140.
This work has received further funding from the Swiss State Secretariat for Education, Research and Innovation (SERI) under contract number MB22.00072, as well as from the Swiss National Science Foundation (SNSF) through project grant 200020\_207349. J.R.W. acknowledges that support for this work was provided by The Brinson Foundation through a Brinson Prize Fellowship grant.
\end{acknowledgements}

%
%

\bibliography{Euclidz6Forcasts} 

%

\begin{appendix}

\section{\label{sc:EF_Paper_Discussion_OtherCrietria} Other selection criteria}
In this section, we discuss other methods of selecting high-$z$ sources using our synthetic catalogues. In Sect. \ref{sc:EF_Paper_Discussion_Others_ext-opt}, we discuss the impacts on the completeness and purity when including additional S/N cuts with the external optical bands (HSC and/or LSST). Sections  \ref{sc:EF_Paper_Discussion_Others_EuclidOnly} and \ref{sc:EF_Paper_Discussion_Others_NoOptExt} explains the impacts of removing ground-based and MIR coverage in SED fitting. In these two sections, the results of the methods are presented as confusion matrices, and we refer the reader to Sect. \ref{sc:EF_Paper_Results_RecoveryPurity} where these matrices are explained.

\subsection{\label{sc:EF_Paper_Discussion_Others_ext-opt} Including ground-based optical data within the signal-to-noise criteria}

As part of the \Euclid mission, the EDFs will contain complementary optical coverage from Subaru/HSC and/or Rubin/LSST \citep{EP-McPartland}. These bands cover a narrower wavelength range compared to \IE, thus can be a useful tool to better sample the Lyman break and distinguish contaminants from high-$z$ sources. 

Due to the Lyman break feature, we do not expect to measure any significant flux measurements in bands bluer than the break. Since the Lyman break of $z>6$ sources falls in the $z$-band, $z>6$ LBGs can be selected by applying S/N cuts in ground-based bands as well as the \Euclid \IE band.
With a $\mathrm{S/N} <2$ in all optical bands, we find that the contamination percentage of low-$z$ sources and MLTs is $0.1\%$ and $36.5\%$, respectively. Although this selection reduces the MLT contamination by $\sim 10\%$ in comparison to the method discussed in Sect. \ref{sc:EF_Paper_Method}, the percentage of high-$z$ sources also reduces by $\sim 10\%$ in each of the EDFs. We find that SED fitting can remove a significant number of MLTs without compensating for the high-$z$ completeness.

\subsection{\label{sc:EF_Paper_Discussion_Others_EuclidOnly}Selecting high-$z$ sources with only \Euclid bands}

The optimal criteria, described in Sect. \ref{sc:EF_Paper_Method}, includes \Euclid, ground-based optical (HSC and/or LSST), and MIR ({\it Spitzer}) bands. In this section, we discuss the impact on the completeness and purity of the final selections after removing ground-based and {\it Spitzer} coverage from the criteria, such that only the \Euclid bands are included.

\begin{figure*}
     \centering
     \begin{subfigure}[b]{0.735\textwidth}
         \centering
         \includegraphics[width=1\textwidth]{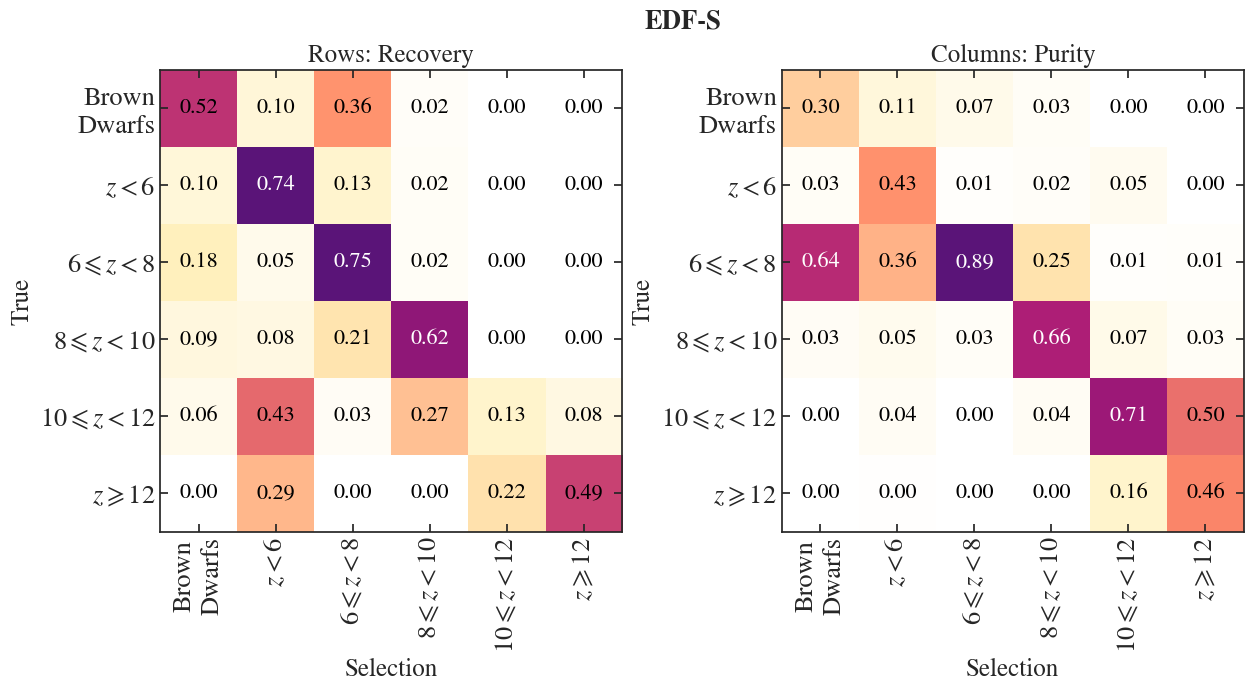}
         \label{fig:recoved_sel_frac_euclidonly_sep}
     \end{subfigure}
     \begin{subfigure}[b]{0.735\textwidth}
         \centering
         \includegraphics[width=1\textwidth]{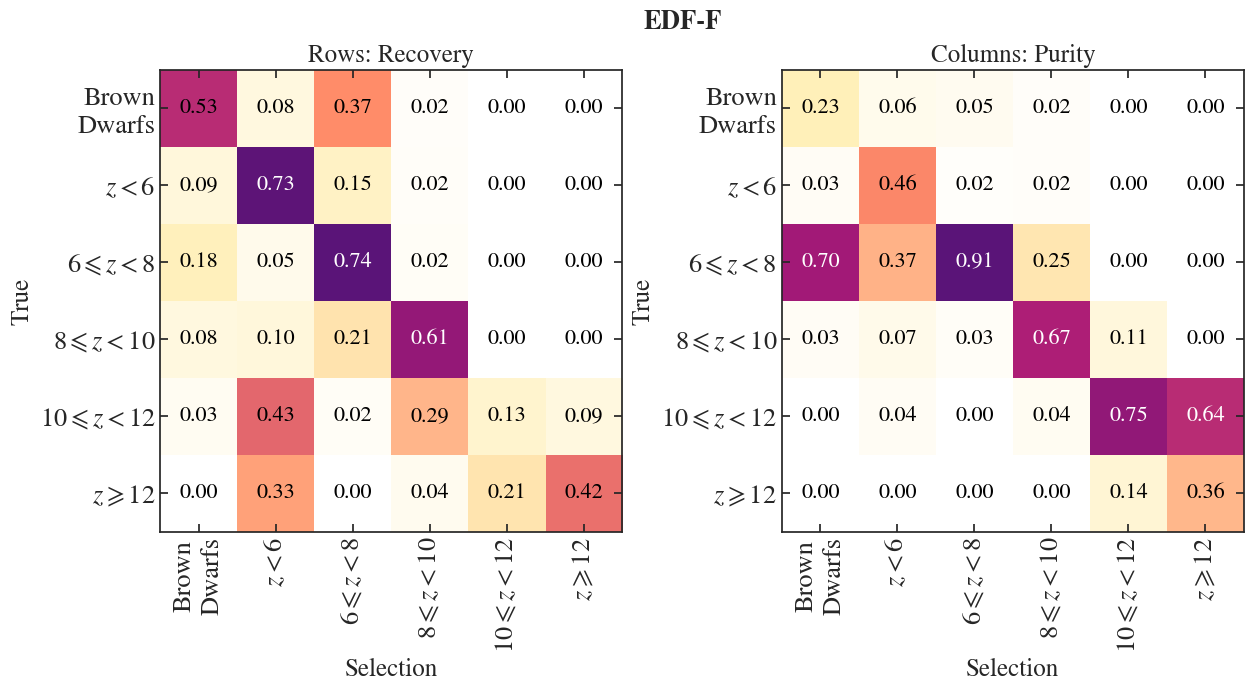}
         \label{fig:recoved_sel_frac_euclidonly_cdfs}
     \end{subfigure}
     \begin{subfigure}[b]{0.735\textwidth}
         \centering
         \includegraphics[width=1\textwidth]{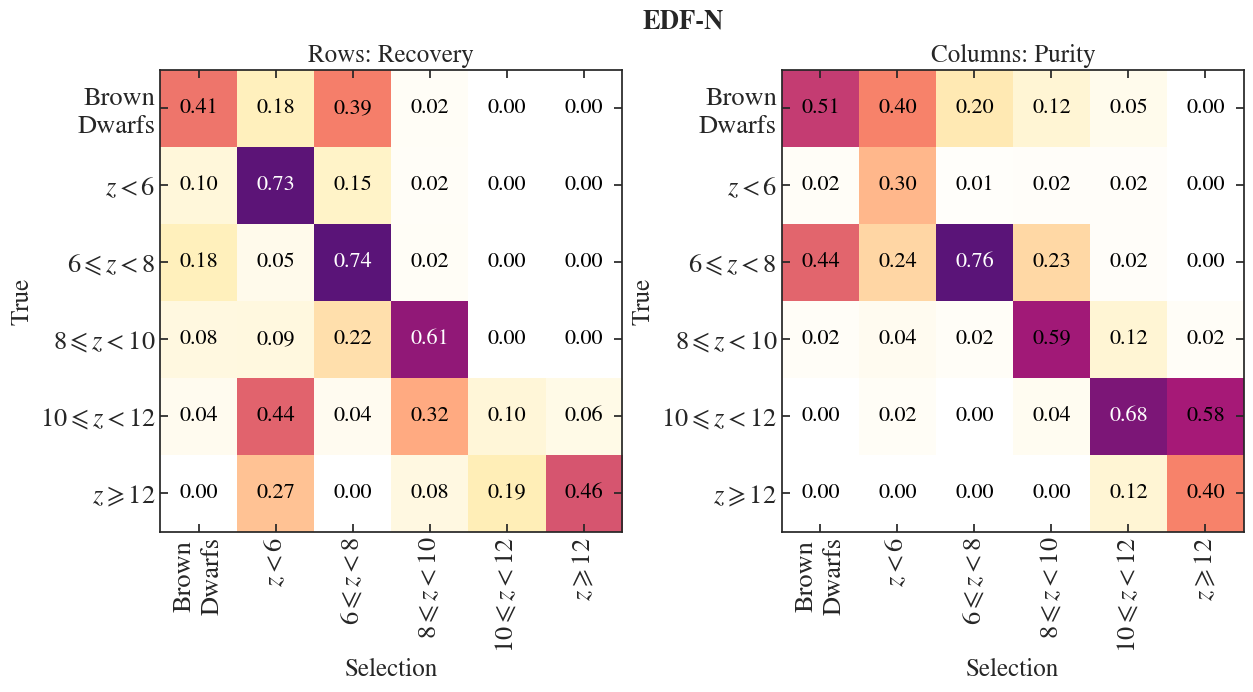}
         \label{fig:recoved_sel_frac_euclidonly_nep}
     \end{subfigure}
     \caption{Performance of a classifier including \Euclid filters only, for each EDF. The performance of this classification is presented in confusion matrices, which are normalised by true class (\textit{left}) and selected class (\textit{right}). Diagonals present completeness (\textit{left}) and purity (\textit{right}).}
        \label{fig:recoved_sel_frac_euclidonly_allfields}
\end{figure*}

We use a similar method to the method used in Sect. \ref{sc:EF_Paper_Method}, but \texttt{eazy} is set to fit templates over the \Euclid bands only. 
The confusion matrices in Fig. \ref{fig:recoved_sel_frac_euclidonly_allfields} present the results of this method for each EDF. 

When comparing the completeness in Fig. \ref{fig:recoved_sel_frac_euclidonly_allfields} to Fig. \ref{fig:recoved_sel_frac_allfields}, we find that removing ground-based optical and {\it Spitzer} MIR coverage reduces the completeness of all classes, for all EDFs. Specifically, the removal of these bands impacts the recovery of $z>10$ sources. For example, the completeness of $10<z<12$ sources in EDF-S is $13\%$ without the complementary data. In each of the left matrices of Fig. \ref{fig:recoved_sel_frac_euclidonly_allfields}, we see a larger fraction ($\sim 43\%$) of $10<z<12$ sources being misclassified as $z<6$ sources. Therefore, without the complementary optical and MIR data, the Lyman break of $z>10$ sources is confused with the Balmer break of $z<6$ sources, causing a reduction in the completeness of $z>10$ sources. 

As shown in the top row of each of the left matrices in Fig. \ref{fig:recoved_sel_frac_euclidonly_allfields}, the recovery of MLTs is $\sim 30$--$40\%$ lower compared to the MLT recovery rates using the optimal method. Without the complementary data,  more MLTs are classified as $6\le z<8$ sources, and this is consistent across all fields. This is seen, for example, along the top row of the top left matrix in Fig. \ref{fig:recoved_sel_frac_euclidonly_allfields}, where $36\%$ of the true MLTs were classified as $6\le z<8$ sources in EDF-S, while only 8\% were placed in this class when external data is included. Thus, the complementary ground-based and space-based MIR coverage is highly important in selecting $z>6$ LBGs with the highest completeness and purity. Without this external coverage, the contamination of MLTs is more significant in $z>6$ selections.

\subsection{\label{sc:EF_Paper_Discussion_Others_NoOptExt}Importance of MIR data in the selection criteria}
In this section, we discuss the effects on the completeness and purity of the final selection when combining \Euclid and {\it Spitzer} MIR coverage in the criteria, but excluding the external ground-based optical data. These results are important for the first sets of \Euclid data releases, as EDF-F and EDF-S will not contain ground-based data at matching depths. 

\begin{figure*}
     \centering
     \begin{subfigure}[b]{0.735\textwidth}
         \centering
         \includegraphics[width=1\textwidth]{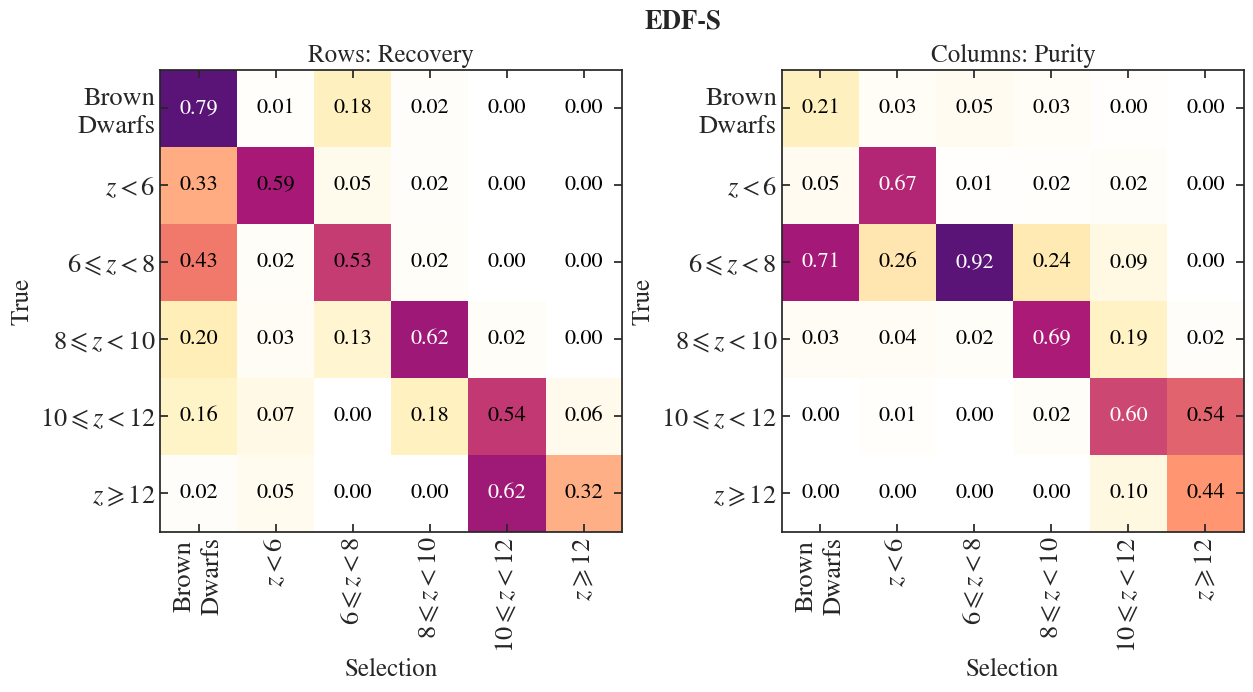}
         \label{fig:recoved_sel_frac_nooptext_sep}
     \end{subfigure}
     \begin{subfigure}[b]{0.735\textwidth}
         \centering
         \includegraphics[width=1\textwidth]{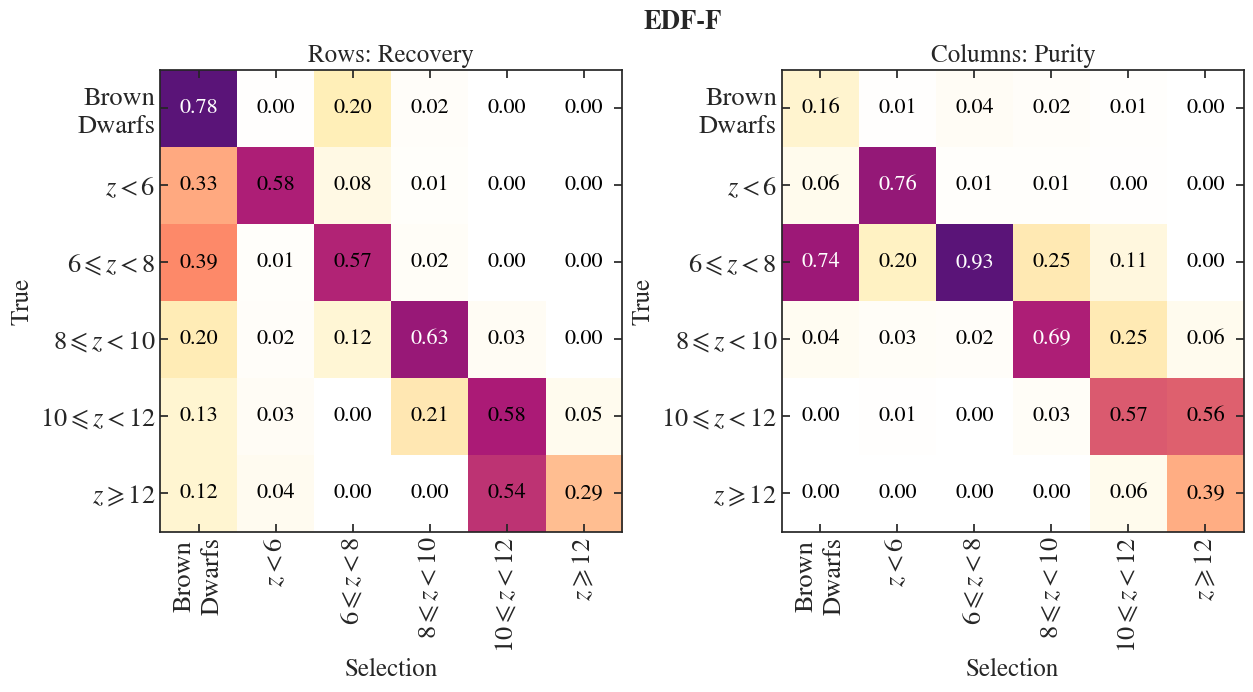}
         \label{fig:recoved_sel_frac_nooptext_cdfs}
     \end{subfigure}
     \begin{subfigure}[b]{0.735\textwidth}
         \centering
         \includegraphics[width=1\textwidth]{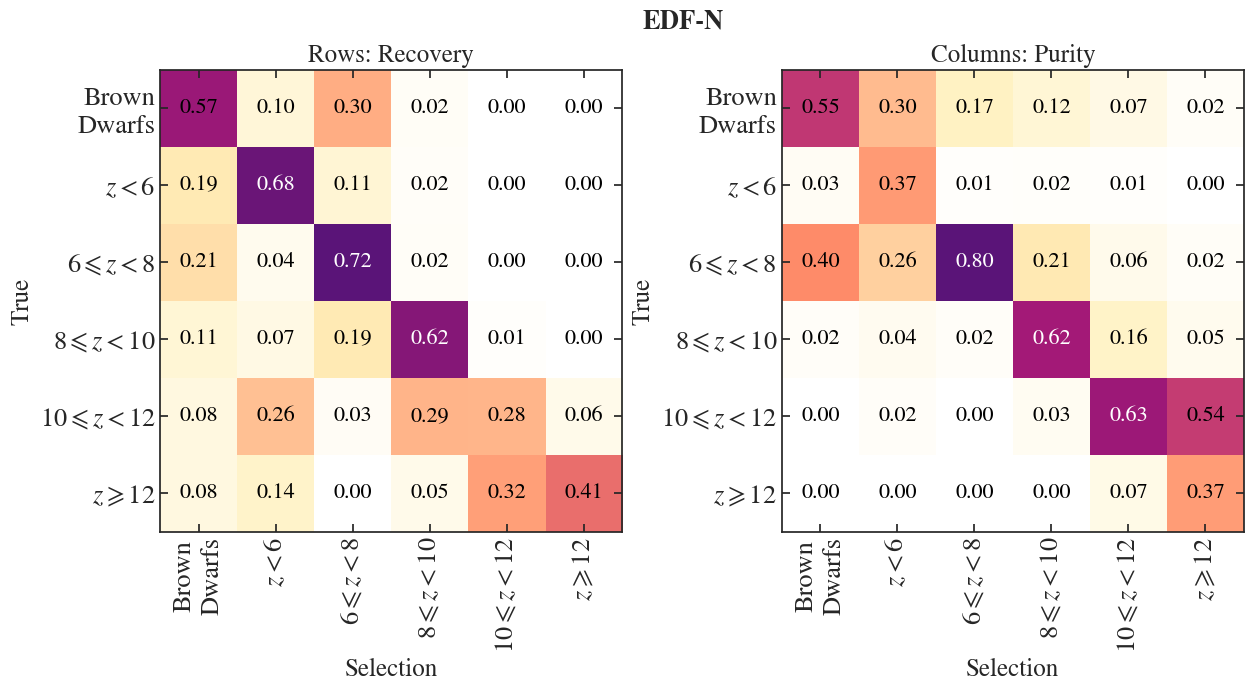}
         \label{fig:recoved_sel_frac_nooptext_nep}
     \end{subfigure}
        \caption{Performance of a classifier which includes both {\it Spitzer} and \Euclid filters, for each EDF. The performance of this classification is presented in confusion matrices, which are normalised by true class (left) and selected class (right). Diagonals present completeness (left) and purity (right).}
        \label{fig:recoved_sel_frac_nooptext_allfields}
\end{figure*}

The final selections are determined using a method similar to the one described in Sect. \ref{sc:EF_Paper_Method}, except that \texttt{eazy} is set to fit templates over \Euclid bands as well as {\it Spitzer}/IRAC Channel 1 and Channel 2 bands.
The results of this method are presented as confusion matrices in Fig. \ref{fig:recoved_sel_frac_nooptext_allfields}, for all EDFs. We refer the reader to Sect. \ref{sc:EF_Paper_Results_RecoveryPurity} for a description of the matrices.

Overall, when including {\it Spitzer}/IRAC data with \Euclid data, the completeness of each class increases. This is shown, for example, in the diagonals of the left matrices in Fig. \ref{fig:recoved_sel_frac_nooptext_allfields}, in comparison to Fig. \ref{fig:recoved_sel_frac_euclidonly_allfields}. With the inclusion of MIR coverage, the recovery of $10\le z<12$ sources increases in all EDFs. For example, the recovery of $10\le z<12$ sources increases to $58\%$ in EDF-F compared to $13\%$, when the {\it Spitzer}/IRAC coverage is not considered in the SED fitting (Fig. \ref{fig:recoved_sel_frac_nooptext_allfields}, middle left matrix). The inclusion of MIR also reduces the number of $z>10$ sources fit as low-$z$ ($z<6$) interlopers. For example, this is shown for EDF-S (Fig. \ref{fig:recoved_sel_frac_nooptext_allfields} top left matrix) where only $7\%$ of true $10\le z<12$ sources are misclassified as $z<6$ sources -- $35\%$ less misclassified sources than a criteria excluded {\it Spitzer} coverage. 

The inclusion of MIR is also important for recovering MLTs and reducing the contamination of MLTs in the $6\le z<8$ selections. For example, in EDF-S, only $18\%$ of the true-MLTs contaminate the $6\le z<8$ sample (Fig. \ref{fig:recoved_sel_frac_nooptext_allfields} top left matrix), in comparison to $36\%$ when MIR is removed (Fig. \ref{fig:recoved_sel_frac_euclidonly_allfields} top left matrix). However, without the inclusion of ground-based optical data, the contamination of MLTs in this class will not drop below $15\%$. Large wavelength coverage from optical to MIR is important to select $z>6$ sources with the highest completeness and purity. 

For EDF-N, the contamination of MLTs in the $6\le z <8$-classified sample is still significantly large when including MIR data in the criteria. The first row of the bottom left matrix in Fig. \ref{fig:recoved_sel_frac_euclidonly_allfields} shows that $30\%$ of true MLTs are still misclassified as $6\le z<8$ sources. This contamination is reduced to $26\%$ when including ground-based optical data. But EDF-N will consistently have the larger contamination levels due to the lack of {\it Spitzer} data in half of the field and because EDF-N lies closer to the Galactic plane, increasing the surface density of MLTs. 

In conclusion, MIR data is essential in the recovery of $z>10$ sources, as well as reducing contamination of MLTs, in all of the EDFs. To further improve the completeness and purity, ground-based data should also be included in the SED fitting process, as this reduces the confusion between Balmer and Lyman breaks of high- and low-z sources.

\subsection{\label{sc:EF_Paper_Discussion_Others_col-col}colour - colour selection criteria}

colour - colour selection can be used to select LBGs as the Lyman break produces a red colour between the bands where the break falls within. We created three colour-colour selection criteria using the colours: ($z-\YE$, $\YE-\JE$), ($\YE-\JE$, $\JE-\HE$), and ($\JE-\HE$, $\HE-{\rm IRAC}_1$), where $ z$ corresponds to $ z_{\rm HSC}$ or $ z_{\rm LSST}$. The colour selection requires that sources have a $\mathrm{S/N}>2$ in the second filter and a $\mathrm{S/N}>5$ in the reddest filter -- unless the reddest filter is a {\it Spitzer}/IRAC band, then we require a $\mathrm{S/N}>5$ in the second filter only. For $z$-dropouts, the selection is as follows
\begin{multline}
    (\mathrm{S/N}_{\YE} > 2) \wedge (\mathrm{S/N}_{\JE} > 5)\wedge \\
    (z - \YE > 1.14 )\wedge (\YE - \JE < 0.5) \wedge (z - \YE > 4\,[\YE- \JE] + 1).
\end{multline}
For \YE-dropouts,

\begin{dmath}
(\mathrm{S/N}_{\JE} > 2) \wedge (\mathrm{S/N}_{\HE} > 5)\wedge \\
    (\YE - \JE > 1 )\wedge (\JE - \HE < 0.5) ,
\end{dmath}
and for \JE-dropouts 

\begin{dmath}
(\mathrm{S/N}_{\HE} > 5) \wedge \\
    (\JE - \HE > 0.99 ) \wedge (\HE - {\rm IRAC}_1 < 1.33) .
\end{dmath}

To further remove contamination in the final sample, we also require S/N and $\chi^2_{\rm opt}$ cuts in the optical coverage (ground-based and \IE). The $\chi^2_{\rm opt}$ cut is derived similarly as \cite{bouwens2011_uvlf}, where we choose the $\chi^2_{\rm opt}$ threshold to remove only $10\%$ of high-$z$ sources, depending on the number of bands used.
With this criteria, we select $\le 20\%$ of the $z=6$--8 ($z$-dropouts) sources from the parent sample. Less conservative colour cuts can improve this recovery rate, but the contamination quickly increases above $15\%$. Including SED fitting from \texttt{eazy} can reduce the contamination to $<20\%$ in all fields, but the incompleteness of $z=6$--8 is still around $20\%$ because of the colour selection.
We also find that the ($\JE-\HE$, $\HE-{\rm IRAC}_1$) colour selects a broad redshift range, from $z=10$ to $z=15$, which makes it difficult to constrain the LF in specific redshift bins. Therefore, based on our synthetic catalogue, we find that colour - colour selection is not required to select a clean sample of $z>6$ LBGs from the final \Euclid data release.

\section{Number counts} 

\subsection{Expected number counts}
This Appendix contains Table \ref{tbl:Q1_expectednumbercounts}, which presents the number counts for high-$z$ sources expected in Q1, in various $z$ bins, for the DPL (models used in the synthetic catalogue) and Schechter (\citealt{whitler25_jades_uvlf}) models. These values are presented visually in Fig. \ref{fig:expnumcounts_hist}. The number counts are calculated by integrating over the UV LF model over an area of 53\,$\rm deg^2$ and down to 24.5\,AB depth. In the first \Euclid dataset, we can already distinguish between the DPL and Schechter models, since we do not expect to detect any sources at $z>10$. 

\begin{table}
\centering
\caption{Expected number counts of $z>6$ galaxies (with Poisson errors)  at the Q1 depth (24.5\,AB) and $53\, \rm deg^2$ area, based on the DPL (\citealt{bowler2020_uvlf_ultravista} at $z<8$ and \citealt{donnan24_jwst_uvlf} at $z\ge 8$) and Schechter (\citealt{whitler25_jades_uvlf}) UV LF.}
\setlength{\tabcolsep}{1.em} 
{\renewcommand{\arraystretch}{1.3}
\begin{tabular}{c|c|c}
Redshift & DPL & Schechter\\
\hline \hline
$6\le z < 8$ & $267 \pm 16$ & $231 \pm 15$ \\
$8\le z < 10$ & $31 \pm 6$ & $2^{+2}_{-1}$ \\
$10\le z < 12$ & $9 \pm 3$ & $0^{+1}_{-0}$ \\
$12\le z < 14$ & $2^{+2}_{-1}$ & $0^{+1}_{-0}$ \\
\end{tabular}}
\label{tbl:Q1_expectednumbercounts}
\end{table}

\subsection{MLT dwarfs}
\begin{figure}
    \centering
    \includegraphics[width=0.8\linewidth]{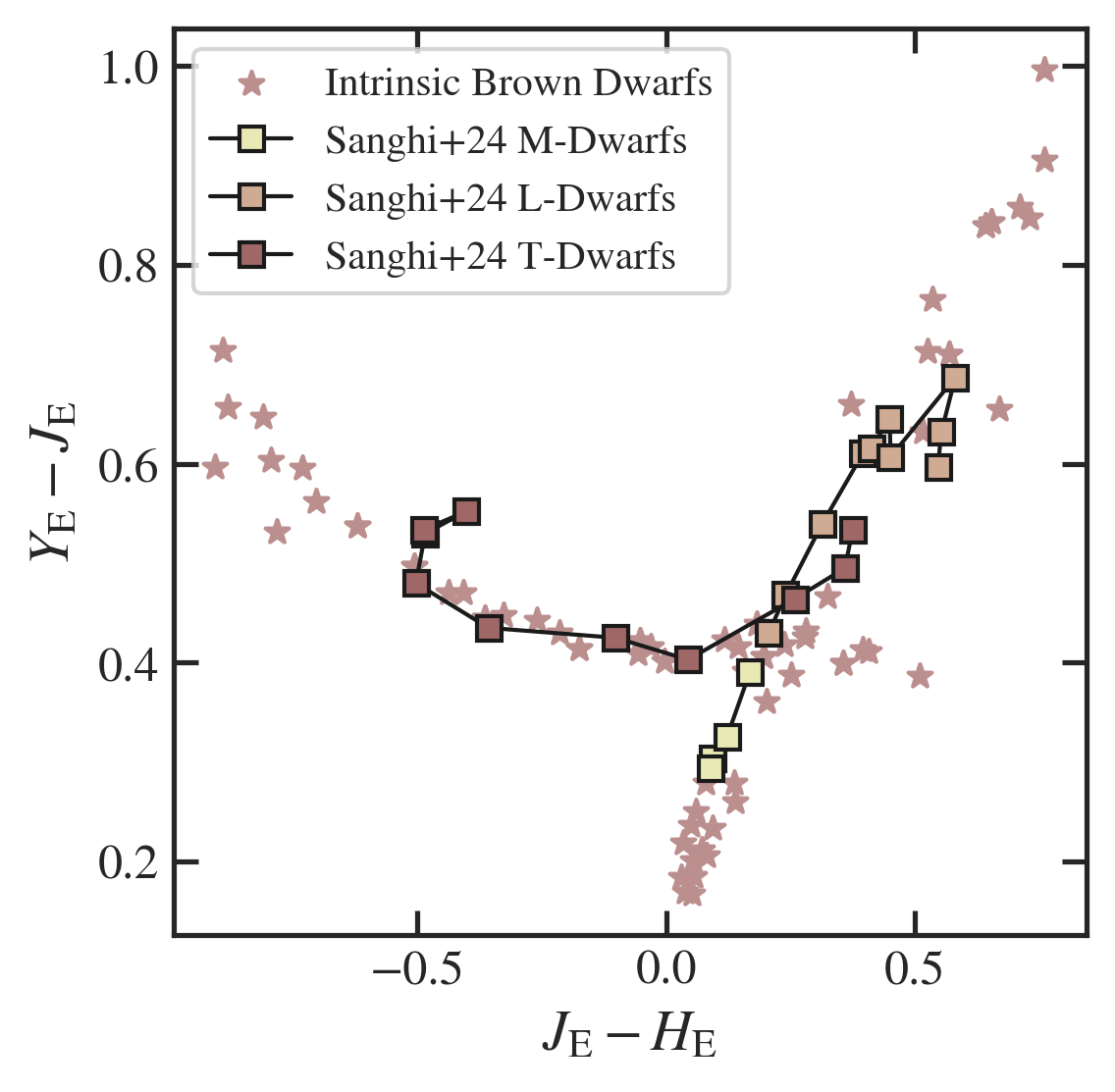}
    \caption{\YE, \JE, \HE colours of the synthetic MLTs in our catalogue presented with the colour-tracks from \citet{sanghi24_bd_color}.}
    \label{fig:bd-colours}
\end{figure}

In this Appendix, we present the $\YE-\JE$ and $\JE-\HE$ colours of the MLTs in the synthetic catalogue in Fig. \ref{fig:bd-colours}. Stars in Fig. \ref{fig:bd-colours} show the intrinsic colours of the synthetic MLTs, and the squares show the interpolated colours from \cite{sanghi24_bd_color}. 



\end{appendix}

\end{document}